%% file: Yousef2018CAD.tex
\def\MYJOURNAL{0} 

\if 0\MYJOURNAL%
\documentclass[3p, preprint, sort&compress, a4paper]{elsarticle}
\usepackage{setspace}
\else \if 1\MYJOURNAL%
\documentclass[journal,10pt,twocolumns]{IEEEtran}
\usepackage{geometry}%
\usepackage[numbers,sort&compress]{natbib}
\fi
\fi

\usepackage{color, xcolor, colortbl}
\usepackage{hyperref}
\hypersetup{
  colorlinks,
  citecolor=blue,
  filecolor=blue,
  linkcolor=blue,
  urlcolor=blue
}
\usepackage{amsfonts}
\usepackage{amsmath}
\usepackage{amssymb}
\usepackage{graphicx}
\usepackage{url}
\usepackage{fp}
\providecommand{\U}[1]{\protect\rule{.1in}{.1in}}
\newtheorem{theorem}{Theorem}

\newtheorem{definition}[theorem]{Definition}

\usepackage{enumitem}
\usepackage[referable,para]{threeparttablex}
\usepackage{footnote}
\usepackage{booktabs}  

\usepackage{mwe}
\usepackage[percent]{overpic}

\usepackage{listings}
\definecolor{dkgreen}{rgb}{0,.6,0}
\definecolor{dkblue}{rgb}{0,0,.6}
\definecolor{dkyellow}{cmyk}{0,0,.8,.3}

\lstdefinestyle{customphp}{
  language        = php,
  basicstyle      = \small\ttfamily,
  keywordstyle    = \color{dkblue},
  stringstyle     = \color{red},
  identifierstyle = \color{dkgreen},
  commentstyle    = \color{gray},
  emph            =[1]{php},
  emphstyle       =[1]\color{black},
  emph            =[2]{if,and,or,else},
  emphstyle       =[2]\color{dkyellow}}

\lstdefinestyle{customc}{
  breaklines=true, breakindent=20pt,
  frame=leftline,
  numbers=left,
  language=C, numberstyle=\tiny, numbersep=10pt,
  showstringspaces=false,
  basicstyle=\footnotesize\ttfamily,
  keywordstyle=\bfseries\color{green!40!black},
  commentstyle=\itshape\color{purple!40!black},
  identifierstyle=\color{blue},
  stringstyle=\color{orange},
  captionpos=t
}

\graphicspath{{./Graphics/}}

\usepackage{fourier}
\DeclareMathAlphabet{\mathcal}{OT1}{pzc}{m}{it}
\DeclareSymbolFont{letters}{OML}{cmm}{m}{it}

\usepackage{tikz}
\usepackage{float}
\usepackage{pgf}
\usepackage{pgfplots}

\usetikzlibrary{calc}
\usepgflibrary{shapes}
\usetikzlibrary{er}
\usetikzlibrary{arrows,snakes,backgrounds}
\usetikzlibrary{decorations.text}

\usepgfplotslibrary{external}
\usetikzlibrary{spy}

\def\getangle(#1) (#2)#3{%
  \begingroup%
  \pgftransformreset%
  \pgfmathanglebetweenpoints{\pgfpointanchor{#1}{center}}{\pgfpointanchor{#2}{center}}%
  \expandafter\xdef\csname angle#3\endcsname{\pgfmathresult}%
  \endgroup%
}

\pgfplotsset{compat=1.11}
\usetikzlibrary{calc,trees,positioning,arrows,chains,shapes.geometric,%
  decorations.pathreplacing,decorations.pathmorphing,shapes,%
  matrix,shapes.symbols}

\tikzset{
  >=stealth',
  punktchain/.style={
    font=\scriptsize,
    rectangle,
    rounded corners,
    draw=black, thick,
    text width=10em,
    minimum height=1em,
    text centered},
  line/.style={draw, thick, <-},
  element/.style={
    tape,
    top color=white,
    bottom color=blue!50!black!60!,
    minimum width=8em,
    draw=blue!40!black!90, very thick,
    text width=10em,
    minimum height=1em,
    text centered},
  every join/.style={->, thick,shorten >=1pt},
  decoration={brace},
  tuborg/.style={decorate},
  tubnode/.style={midway, right=2pt},
}

\if 0\MYJOURNAL%
\tikzset{
  PIXEL/.style={
    font=\fontsize{4}{3.6}\selectfont,
    text width=9em,
    minimum height=1em,
    text centered
  }
}
\else \if 1\MYJOURNAL%
\tikzset{
  PIXEL/.style={
    font=\tiny,
    text width=8em,
    minimum height=3em,
    text centered
  }
}
\fi
\fi

\newcommand{\TMPSPACE}{1in} 

\newcommand{\MYCAD}{LIBCAD}

\usepackage{todonotes}
\usepackage{changebar}

\begin{document}

\if 0\MYJOURNAL%
\input{SecAbstract-PR.tex}
\else \if 1\MYJOURNAL%
\input{SecAbstract-TMI.tex}
\fi
\fi

\input{SecIntroduction.tex}

\input{SecBackground.tex}

\input{SecAlgorithms.tex}

\input{SecAssessment.tex}

\input{SecSystem.tex}

\input{SecDiscussion.tex}

\input{SecAcknowledgment.tex}

\section{Appendix}\label{sec:appendix}
\input{Algorithms.tex}

\if 0\MYJOURNAL%
\bibliographystyle{elsarticle-num}
\bibliography{publications,booksIhave}
\else \if 1\MYJOURNAL%
\small
\section*{References}
\bibliographystyle{IEEEtran}
\bibliography{publications,booksIhave}
\normalsize
\fi
\fi

\input{Figures.tex}

\end{document}

%% file: SecAbstract-PR.tex
\begin{frontmatter}
  \title{Method and System for Image Analysis to Detect Cancer\tnoteref{t1}} \tnotetext[t1]{This
    title is chosen to be the same as the title of the pending patent of this
    project~\citep{Yousef2017MethodSystemForComputer}. This project was funded collaboratively by
    (1) \textit{Information Technology Industry Development Agency} (ITIDA
    \url{http://www.itida.gov.eg}), grant number ARP2009.R6.3.; (2) \textit{MESC for Research and
      Development} (MESC Labs \url{www.mesclabs.com})}

  \author[WAY]{Waleed~A.~Yousef\corref{cor1}}
  \ead{wyosuef@GWU.edu, wyousef@fci.helwan.edu.eg}
  \cortext[cor1]{Corresponding Author}

  \author[AAA]{Ahmed A. Abouelkahire\fnref{fn1}}
  \ead{ahmedanis03@gmail.com}

  \author[DEA]{Deyaaeldeen~Almahallawi\fnref{fn1}}
  \ead{DAlmahal@indiana.edu}

  \author[OSM]{Omar~S.Marzouk\fnref{fn1}}
  \ead{omares@kth.se}

  \author[SKM]{Sameh~K.\ Mohamed\fnref{fn1}}
  \ead{sameh.kamal@insight-centre.org}

  \author[WAM]{Waleed~A.~Mustafa\fnref{fn1}}
  \ead{5mustafa@informatik.uni-hamburg.de}

  \author[OMS]{Omar~M.~Osama\fnref{fn1}}
  \ead{omar.m.osama1989@gmail.com}

  \author[AAS]{Ali~A.~Saleh\fnref{fn1}}
  \ead{3saleh@informatik.uni-hamburg.de}

  \author[NMA]{Naglaa~M.~Abdelrazek}
  \ead{NaglaAbdelrazek@Yahoo.com}

  \fntext[fn1]{These authors contributed equally to the manuscript, their names are ordered
    alphabetically according to the family name, and each of them is the second author.}

  \address[WAY]{Ph.D., Computer Science Department, Faculty of Computers and Information, Helwan
    University, Egypt.\\ Human Computer Interaction Laboratory (HCI Lab.), Egypt.}

  \address[AAA]{B.Sc., Senior Data Scientist, TeraData, Egypt.}

  \address[DEA]{M.Sc., School of Informatics and Computing, Indiana University Bloomington, USA.}

  \address[OSM]{M.Sc., School of Computer Science and Communication (CSC), KTH Royal Institute of
    Technology, Sweden.}

  \address[SKM]{M.Sc., Insight Center For Data Analytics, National University of Ireland, Ireland.}

  \address[WAM]{B.Sc., Department of Informatics, University of Hamburg, Germany.}

  \address[OMS]{B.Sc., MESC for Research and Development, Egypt.}

  \address[AAS]{M.Sc., Hamburg University, Germany.}

  \address[NMA]{Ph.D., M.D., Professor, Faculty of Medicine, Cairo University; Quality Control
    Manager, Alfa Scan Radiology Center; and Quality Control Manager At National Screening
    Campaign, Egypt}

  \begin{abstract}
    \input{abstract.tex}
  \end{abstract}

  \begin{keyword}
    \input{keywords.tex}
  \end{keyword}

\end{frontmatter}


%% file: abstract.tex
Breast cancer is the most common cancer and is the leading cause of cancer death among women
worldwide. Detection of breast cancer, while it is still small and confined to the breast, provides
the best chance of effective treatment. Computer Aided Detection (CAD) systems that detect cancer
from mammograms will help in reducing the human errors that lead to missing breast
carcinoma. Literature is rich of scientific papers for methods of CAD design, yet with no complete
system architecture to deploy those methods. On the other hand, commercial CADs are developed and
deployed only to vendors' mammography machines with no availability to public access. This paper
presents a complete CAD; it is complete since it combines, on a hand, the rigor of algorithm design
and assessment (method), and, on the other hand, the implementation and deployment of a system
architecture for public accessibility (system). (1) We develop a novel algorithm for image
enhancement so that mammograms acquired from any digital mammography machine look qualitatively of
the same clarity to radiologists' inspection; and is quantitatively standardized for the detection
algorithms. (2) We develop novel algorithms for masses and microcalcifications detection with
accuracy superior to both literature results and the majority of approved commercial systems. (3) We
design, implement, and deploy a system architecture that is computationally effective to allow for
deploying these algorithms to cloud for public access.



%% file: keywords.tex
Breast Cancer\sep Detection\sep Mammography\sep Mammograms\sep Image Processing\sep Pattern
Recognition\sep Computer Aided Detection\sep CAD\sep Classification\sep Assessment\sep FROC.


%% file: SecAbstract-TMI.tex
\title{Method and System for Image Analysis to Detect Cancer}

\author{%
  Waleed~A.~Yousef,~\IEEEmembership{Senior Member,~IEEE,}~\thanks{Waleed A. Yousef is an associate
    professor, Human Computer Interaction Laboratory (HCI Lab.), computer science department,
    faculty of computers and information, Helwan University, Egypt,
    \url{wyousef@fci.helwan.edu.eg}} %
  Ahmed A. Abouelkahire\textsuperscript{a}~\thanks{Ahmed A. Abouelkahire, B.Sc., Senior Data Scientist, TeraData, Egypt,
    \url{ahmedanis03@gmail.com}}
  Deyaaeldeen~Almahallawi\textsuperscript{a},~\thanks{Deyaaeldeen Almahallawi, M.Sc., School of Informatics and Computing,
    Indiana University Bloomington, USA, \url{DAlmahal@indiana.edu}} %
  Omar~S.~Marzouk\textsuperscript{a},~\thanks{Omar S. Marzouk, M.Sc., School of Computer Science and Communication (CSC), KTH
    Royal Institute of Technology, Sweden, \url{omares@kth.se}} %
  Sameh~K.\ Mohamed\textsuperscript{a},~\thanks{Sameh K. Mohamed, M.Sc., Insight Center for Data Analytics, National
    University of Ireland, Ireland, \url{sameh.kamal@insight-centre.org}} %
  Waleed~A.~Mustafa\textsuperscript{a},~\thanks{Waleed A. Mustafa, B.Sc., Department of Informatics, University of
    Hamburg, Germany, \url{WaleedAMustafa@gmail.com}} %
  Omar~M.~Osama\textsuperscript{a},~\thanks{Omar M. Osama, B.Sc., MESC Labs., Egypt,
    \url{omar.m.osama1989@gmail.com}} %
  Ali~A.~Saleh\textsuperscript{a},~\thanks{Ali A. Saleh, M.Sc., Hamburg University, Germany, \url{ali.saleh@fcih.net}} %
  Naglaa~M.~Abdelrazek~\thanks{Naglaa M. Abdelrazek, Ph.D., M.D., is a professor, faculty of
    medicine, Cairo University, Egypt, quality control manager at Alfa Scan radiology center, and
    quality control manager at national screening campaign, \url{NaglaAbdelrazek@Yahoo.com}} %

  \thanks{\textsuperscript{a}These authors contributed equally to the manuscript, their names
    are ordered alphabetically according to the family name, and each of them is the second author.}
  \thanks{This project has a patent pending status~\citep{Yousef2017MethodSystemForComputer}. This
    project was funded collaboratively by: (1) ITIDA (Information Technology Industry Development
    Agency), \url{http://www.itida.gov.eg}, grant number ARP2009.R6.3.; (2) MESC Labs for Research
    and Development \url{www.mesclabs.com}.}
}

\maketitle

\begin{abstract}
  \input{abstract.tex}
\end{abstract}

\newcommand{\sep}{,~} 
\begin{IEEEkeywords}
  \input{keywords.tex}
\end{IEEEkeywords}


%% file: SecIntroduction.tex
\section{Introduction}\label{sec:introduction}

\subsection{Breast Cancer and Computer Aided Detection (CAD)}\label{sec:breast-canc-comp}
Breast cancer is the most common cancer in women in developed Western countries and is becoming ever
more significant in many developing countries~\citep{Althuis2005GlobalTrendsBreastCancer}. Although
incidence rates are increasing, mortality rates are stable, representing an improved survival
rate. This improvement can be attributed to effective means of early detection, mainly mammography,
as well as to significant improvement in treatment
options~\citep{Freedman2006CancerIncidenceMECC}. Moreover, ``\textit{Breast carcinoma is not a
  systemic disease at its inception, but is a progressive disease and its development can be
  arrested by screening}'' \citep{Tabar1999NaturalHistCarcinoma,
  Anderson2000NaturalHistCarcinoma}. Early detection of breast cancer, while it is still small and
confined to the breast, provides the best chance of effective treatment for women with the
disease~\citep{Ferlay2004CancerIncidence,Parkin2005GlobalCancer}.

Detection of breast cancer on mammography is performed by a human reader, usually radiologist or
oncologist. Causes of missed breast cancer on mammography can be secondary to many factors including
those related to the patient (whether inherent or acquired), the nature of the malignant mass
itself, poor mammography techniques, or interpretive skills of the human reader (including
perception and interpretation errors)~\citep{Kamal2007MissedBreast}. Perception error occurs when
the lesion is included in the field of view and is evident but is not recognized by the
radiologist. The lesion may or may not have subtle features of malignancy that cause it to be less
visible. Small nonspiculated masses, areas of architectural distortion, asymmetry, small clusters of
amorphous, or faint microcalcifications, all may be difficult to
perceive~\citep{Kamal2007MissedBreast}. On the other hand, interpretation error occurs for several
reasons including lack of experience, fatigue, or inattention. It may also occur if the radiologist
fails to obtain all the views needed to assess the characteristics of a lesion or if the lesion is
slow growing and prior images are not used for
comparison~\citep{Radhika2005ImagingTechniques,Kamal2007MissedBreast}.

To aid human readers and to minimize the effect of both the perception and interpretation errors,
CADs were introduced. A Computer Aided Detection (or Computer-Assist Device) (CAD) ``\textit{refers
  to a computer algorithm that is used in combination with an imaging system as an aid to the human
  reader of images for the purpose of detecting or classifying
  disease}''~\citep{Wagner2002AssessmentOf}. CAD systems reduces the human errors that lead to
missing breast carcinoma, either related to poor perception or interpretation errors, which could
increase the sensitivity of mammography interpretation~\citep{Muttarak2006BreastCarcinomas}. %
CAD in US receives attention from radiologists and its use increased rapidly; however, this is not
the case in many parts of the world~\citep{RaoHowWidelyCAD2010}.

\bigskip

In retrospect, it is of high importance to the field of medical diagnosis to design a complete CAD
that: (1) works on mammograms acquired from any mammography machine; (2) posses highest possible
sensitivity at fairly low specificity, (3) deploys to cloud for public accessibility and is not
explicit to a particular in-site mammography machine. To fulfill these three objectives, we launched
this project.

\subsection{Project: Objectives, Protocol, and Database}\label{sec:proj-prot-data}
To fulfill the three objectives just mentioned above, we launched this project in which we were able
to design a novel method (a set of algorithms) and a novel system (software architecture) detailed
in this article. Some results of early stages of this project had been published
in~\cite{Yousef2010OnDetecting, AbdelRazek2012MicroclacificationLIBCAD,
  AbdelRazek2013MicroclacificationLIBCAD}. The present article is the main publication of the
project, and is almost self-contained, where all algorithms and experiments are detailed. We
postponed publishing this article until the conclusion of the project and filing the
patent~\citep{Jalalian2017FoundationMethodologiesComputerAided}.

The working team of this project comprises a multidisciplinary group of several backgrounds
including statistics, computer science, and engineering, along with a trained, experienced and
professional radiologist (10 years experience, 5000 mammogram / year). We collected digital
mammograms from two different institutions, three different mammography machines, with different
resolutions. This is to test on testing datasets totally different from training datasets to confirm
CAD generalization (Sec.~\ref{sec:different-versions}). The description and the main properties of
these different datasets are explained in Table~\ref{tab:datab-acqu-design}.
\begin{table}[!tb]
  \centering
  \begin{tabular}[t]{llllll}
    \toprule
                          & \textbf{D1}           & \textbf{D2}           & \textbf{D3}      & \textbf{D4}        & \textbf{D5}           \\
    \midrule
    \textbf{\# cases}     & 55                    & 81                    & 1149             & 88                 & 99                    \\
    \textbf{\# normal}    & 0                     & 0                     & 663              & 62                 & 99                    \\
    \textbf{\# malignant} & 55                    & 81                    & 60               & 4                  & 0                     \\
    \textbf{\# benign}    & 0                     & 0                     & 426              & 22                 & 0                     \\
    \textbf{Image Width}  & 1914                  & 1914                  & 4728             & 2364               & 1914                  \\
    \textbf{Image Height} & 2294                  & 2294                  & 5928             & 2964               & 2294                  \\
    \textbf{BitDepth}     & 12                    & 12                    & 16               & 10                 & 12                    \\
    \textbf{Modality}     & FFDM                  & FFDM                  & FFDM             & CR                 & FFDM                  \\
    \textbf{Manufacturer} & GE                    & GE                    & FUJIFILM         & FUJIFILM           & GE                    \\
    \textbf{Institution}  & A                     & A                     & B                & B                  & A                     \\
    \bottomrule
  \end{tabular}
  \caption{Properties and attributes of five databases acquired for design and assessment of CAD
    algorithms. The two institutions A and B are designates the national campaign for breast cancer
    screening and Alfa Scan private radiology center. The two datasets D4 and D5 are used for
    measuring the false positive since the majority of them are normal cases
    (Sec.~\ref{sec:different-versions})}\label{tab:datab-acqu-design}
\end{table}

We have implemented our protocol according to which the experienced radiologist read the digital
mammograms and then marked and attributed lesions on images with the aids of a software developed in
the project for that purpose; Figure~\ref{fig:snapsh-our-design} is a snapshot of this
software. Radiologist's annotations are done manually without semi-automation of any active contour
learning algorithm~\citep[e.g.,][]{Zhao2018MinimizationAnnotationWork,
  Zhao2018MammographicImageClassificationSystem}. The marked lesions are then tagged according to
the different radiology lexicons and then categorized by the radiologist according to the
``\textit{Breast Imaging Reporting and Data System}'' (BIRADS) scoring system
(Table~\ref{TableBIRAD}). All suspicious lesions tagged by the radiologists as BIRADS 3, 4 or 5 are
pathologically proven after core and vacuum needle biopsy.

\begin{table}[!tb]
  \centering
  \begin{tabular}{cl}
    \toprule
    \textbf{Category} & \textbf{Indication}                   \\
    \midrule
    0                 & mammographic assessment is incomplete \\
    1                 & negative                              \\
    2                 & benign finding                        \\
    3                 & probably benign finding               \\
    4                 & suspicious abnormality                \\
    5                 & highly suggestive of malignancy       \\
    \bottomrule
  \end{tabular}
  \caption{Breast Imaging Reporting and Data System (BIRADS)}\label{TableBIRAD}
\end{table}

\subsection{Organization of Manuscript}\label{sec:organ-manuscr}
The sequel of this manuscript is organized as follows. Section~\ref{sec:review} is a background and
a brief literature review on CAD design and assessment. It reviews the main five steps of CAD
design: breast segmentation, image enhancement, mass detection, microcalcifications detection
(Sec.~\ref{sec:background-segmentation}--\ref{sec:background-micr-detect} respectively), which are
the design steps, and assessment
(Sec.~\ref{sec:background-assessment}). Section~\ref{sec:algorithms} details the design steps of our
method in four subsections parallel to
Sec.~\ref{sec:background-segmentation}--\ref{sec:background-micr-detect}
respectively. Section~\ref{sec:assessment} details the assessment of our method, where the method
accuracy is reported and compared to both the literature methods and commercial
CADs. Section~\ref{sec:deploying-cad-as} details the system design and assessment, where a system
architecture and implementation are proposed and system performance is
measured. Section~\ref{sec:discussion} concludes the article and discusses current and future work
to complement the present article and implemented system.


%% file: SecBackground.tex
\section{Background and Related Work}\label{sec:review}
Designing a CAD to detect cancer from digital mammography involves many steps, including: breast
segmentation (or boundary detection), image enhancement (or filtering), mass detection (or mass
segmentation), microcalcification detection, and finally assessment. Each of these five steps has
its own literature. There are good review papers compiling this literature. E.g.,
\cite{Cheng2003CADsurveyMC, Cheng2006ApproachesAutoDetClass, Bozek2009SurveyImageProc,
  Tang2009CADreview} provide an overview of recent advances, and a survey of image processing
algorithms, used in each of the CAD design steps. %
\cite{Rangayyan2007ReviewofCAD} present some of their recent works on the development of image
processing and pattern analysis techniques. \cite{Oliver2010ReviewMassDet} is a seminal work, and
from our point view is one of the best review papers written in this field, in particular, and in
many other fields, in general. The authors provide both a review of the algorithms of the literature
and a unique comparative study of these algorithms by reproducing them and assessing them on the
same dataset and benchmark. \cite{Adler1995NewMethodsForImaging, Giger1996b, Vyborny2000CAD,
  Singh2008b, Giger2008b, Elter2009CADx} provide an overall review on CAD and its history from the
point of view of radiology and medicine community. It is not the objective of the present paper to
provide a detailed review of the field of CAD design. However, since the objective of this project
is to design a complete CAD system, and hence design each of these individual five steps, a brief
literature review of each of them is due in
Sec.~\ref{sec:background-segmentation}--\ref{sec:background-assessment} respectively.

\subsection{Segmentation}\label{sec:background-segmentation}
Segmentation is the first step in CAD design, where the pixels of the breast are identified from the
pixels of the background. It is called, as well, boundary detection, and should not be confused with
mass segmentation and detection. The task of segmenting the breast from the background of the
mammogram is relatively quite easy for that the breast pixels are spatially isolated from the
background, and for that their gray levels are on the upper range of the gray level scale than the
pixels of the background (of course with probable some overlap of their two histograms). The
majority of the literature of segmentation use histogram based methods with different variants. The
simplest of those methods is global histogram separation based on Otsu's
method~\citep{Otsu1979ThresholdSelectionMethodFrom}. \cite{Ojala2001AccurateSegmentation,
  Ferrari2004IdentificationBreastBoundaryMammograms} provide a good comparison among different
histogram based methods and propose a more elaborate combination between histogram based methods on
a hand and active contour and morphological operations on the other hand. We developed a new
histogram based iterative algorithm, with Otsu's method as its initial seed
(Sec.~\ref{sec:segmentation}).

\subsection{Image Enhancement}\label{sec:background-image-enhancement}
Image enhancement, or noise removal and filtering, is one of the preprocessing steps before
abnormality detection. Since the quality of mammography machines and their settings along with the
noise associated with the mammography process differ from one setup to others, image enhancement
accounts for standardizing/normalizing images as a preprocessing step before the detection
algorithms. \cite{Sivaramakrishna2000ComparingPerformance, Singh2005AnEvaluation} provide some
general survey and comparative study for the different enhancement
algorithms. \cite{Sakellaropoulos2003WaveletBasedSpatially, Scharcanski2006DenoisiningEnhancing,
  Tang2007ImageEnhancementAlgorithm} perform image enhancement in the wavelet
space. \cite{Dhawan1986EnhancementMammoFeatures, Morrow1992RegionBasedContrast,
  Petrick1995Automated, Petrick1996AdaptiveDensityWeighted, Rangayyan1997ImprovementSensitivity,
  Kim1997Adaptivemammographicimage} perform image enhancement by contrasting the gray level of a
pixel with its surrounding, and accordingly modifying it, e.g., by subtracting the noise from the
breast tissue (the signal). They used variant methods to implement this idea including: Adapted
Neighborhood Contrast Enhancement (ANCE), Density Weighting Contrast Enhancement (DWCE), adaptive
image enhancement using first derivative and local statistics, and region-based growing
methods. \cite{Veldkamp2000NormalizationMammo} performs local contrast enhancement by noise modeling
and removal. \cite{Gupta1995TheUseofTexture} introduced a smart idea of standardizing images by
Global Histogram Equalization (GHE), where the gray levels of the breast regions will be enforced to
follow uniform distribution (this is our point of departure for introducing our novel method of
image enhancement (Sec.~\ref{sec:image-enhancement})). \cite{McLoughlin2004NoiseEqualization}
proposed a smart low-level enhancement procedure by modeling the dominating quantum noise of the
X-ray procedure as a function of gray level, then correcting the gray level. They demonstrated the
efficacy of this approach in detection of microcalcification.

\subsection{Mass Detection}\label{sec:background-mass-detection}
A ``Mass'' refers to any circumscribed lump in the breast, which may be benign or malignant. It is
the most frequent sign of breast cancer. Many approaches exist in the literature for mass
detection/segmentation. A simple binary taxonomy of these approaches that helps introducing our
method is: region-based vs. pixel-based. However, the two approaches may overlap and some methods
can subscribe to either approaches.

In region-based approach, a whole region (or object) in the image is either clustered (segmented),
given a probability (score) corresponding to the likelihood of malignancy, or assigned a binary
decision (marker). This is in contrast to the pixel-based approach, where each pixel is first given
a score to constitute a whole score image, which is in turn post-processed. The output of the
post-processing could be, as well, clustered pixels representing regions, a final score for each
region, or a decision marker.

\bigskip

It is worth mentioning that, the Convolution Neural Network (CNN) approach, which allows for
building the recent powerful Deep Neural Networks (DNN), subscribes to the pixel-based
approach. This is sense the whole image is fed directly to the network, which produces a final score
image. In addition, the CNN is a featureless approach, where the CNN works directly on the plain
gray level of the pixels without handcrafting any features or algorithms. Even the boundary
detection (Sec.~\ref{sec:background-segmentation}) comes as a direct byproduct of the network
learning process and the final detection task.

\subsubsection{Region-based approach}\label{sec:regi-based-appr}
\cite{Petrick1995Automated, Petrick1996Automated, Petrick1999CombAdap} use region growing for mass
segmentation. \cite{Ciecholewski2017MalignantBenignMassSegmentation} perform mass segmentation by
active contour method. \cite{Rangayyan1997MeasuresAcut, Sahiner2001Improvement} extracts
morphological feature measures of segmented masses to classify them and to improve detection
accuracy.

Many other articles tried to extract useful features from mammograms by leveraging the literature of
texture analysis (\cite{Ojala1996ComparativeStudy} is an early comparative study of this literature,
aside from application to mammograms). \cite{Harwood1995TextureClassification,
  Gupta1995TheUseofTexture} extract features using Law's texture filter for gradients, and other
higher covariance structure filters. \cite{Chan1995CAD, Sahiner1998Computerized} introduced their
Rubber Band Straightening Transorm (RBST) that straighten (flatten) an image object to a rectangular
region. Then, for the segmented mass, they constructed Spatial Gray Level Dependence (SGLD) matirx,
a form of co-occurrence matrix, which describes how different values of neighbor pixels
co-occur. Different variants of co-occurrence matrix exist depending on the order of statistics used
in analyzing pixel values. Later, \cite{Karssemeijer1996Detection} introduced a smart idea to
construct a map of pixel orientation that reveals probable stellate mass
structure. \cite{Kobatake1999ConvergenceIndx, Kobatake1999CompDetec} elaborated on the idea and
suggested their IRIS filter that extracts gradient and directional information. It is one of the
most successful methods in literature of mass detection and outperforms all other methods that was
reproduced in the benchmark of \cite{Oliver2010ReviewMassDet}. \cite{Varela2007Computerized}
extracted a combination of texture based features including IRIS filter, feature from co-occurrence
matrix variants, and contour related features. \cite{Midya2018EdgeWeightedLocalTexture} introduce a
newer version of the traditional Local Binary Pattern (LBP) co-occurrence
matrix. \cite{Shastri2018DensityWiseTwoStage} extract texture features using Histogram Gradient and
Gabor filter.

Other miscellaneous approaches exist; e.g., \cite{Zwiggelaar1999ModelBased,
  Zwiggelaar1999DetectionMass, Zwiggelaar2004LinearStruc} detect speculated masses by building a
model for the linear irregular structure of the boundary of these
masses. \cite{Hastie1999StatMeasureCAD} attacked the problem of mass detection differently; they do
least square regression for a broken line model to regress the gray level intensity of pixels
surrounding a particular POI (response) on the radius from that POI
(predictor). \cite{Li1997DigitalMammography, Qian1999ImageFeature, Qian2001DigitalMammography,
  Li2002ComputerAided, Rashed2007Multiresolution} extract features from multi-resolution
decomposition based on wavelet analysis. \cite{Anitha2017DualStageAdaptiveThresholding,
  Chakraborty2018ComputerAidedDetectionDiagnosis, Chakraborty2018ComputerAidedDetectionMammographic}
use variants of Gray Level Thresholding (GLT) to suppress pixels with low probability of
malignancy. This inclues, high-to-low intensity thresholding or Dual Stage Adaptive Thresholding
(DSAT). Usually thresholding is combined with other morphological procedures to smooth segmented
regions resulted from thresholding.

\subsubsection{Pixel-based approach}\label{sec:pixel-based-approach}
\cite{Lo1995Artificialconvolution, Sahiner1996ClassMass} are CNN recent attempts before the era of
DNN. \cite{Lai1989OnTechniques, Brake1999SingleMultiDetMammo} use template matching to match the
surrounding region of a pixel with a predefined template of gray level distribution that ``almost''
has the characteristic of mass structure; then assign a matching score to this
pixel. \cite{Freixenet2008Eigendetection} learn and design the gray level distribution of templates
from real masses. \cite{Kegelmeyer1994CompAidMammo} was one of the first attempts in the literature,
if not the first, to work directly at the pixel gray level with no feature handcrafting (featureless
approach). The features of each pixel are the plain gray levels in the region surrounding that
pixel. Then, they train a Classification and Regression Trees
(CART). \cite{Campanini2004Featureless} followed the same procedure using SVM rather than CART, with
modification to the postprocessing steps; (we will follow the same featureless approcah, as well,
but with novel procedure (Sec.~\ref{sec:mass-detection}).)

Ensemble classifier, or Multi Classifier System, almost boosted results in many
fields. \cite{Choi2014CADDetectionEnsemble} leveraged the idea of MCS and combined several mass
detection algorithms to boost the detection accuracy of their final system; (we will leverage MCS as
well (Sec.~\ref{sec:multi-class-syst}))



\subsection{Microcalcification Detection}\label{sec:background-micr-detect}
Clusters of microcalciﬁcations are an early sign of possible cancer and are in general not
palpable. Small clusters of amorphous or faint microcalciﬁcations may be difficult to perceive. From
the results of both literature and industry, it seems that microcalcification detection is much
easier for CADs than mass detection.

\subsubsection{Region-based}\label{sec:region-based}
\cite{Qian1993TreeStruc, Yoshida1995OptimizingWavelet, Strickland1996Wavelet, Chen1997OnDigital,
  Yu2000CADsystem, Heinlein2006IntegratedWavelets, Karahaliou2008BreastCancerDiag} extract features
mainly from wavelet transform. \cite{Singh2006SVMbased} detect microcalcifications using
morphological operations (mainly, Sobel and Canny edge detection); then they extract features from
detected objects, and feed them to SVM classifier to decrease false
positives. \cite{Nishikawa1995CAD, Ge2008VascularCalcification} relies mainly on global thresholding
of image histogram then they cluster the objects using either morphological operations, e.g.,
erosion, or $K$-means clustering. \cite{Karssemeijer1992StochasticModel} build a stochastic model
using random fields to model the microcalcification.

Many articles gave attention to the fact that microcalcifications are very little calcium deposits,
and hence are represented by high frequency components with respect to other background
tissues. \cite{Chan1987ImageFeatureAnaMicrocalc} design two filters, one for signal
(microcalcification of high frequency) enhancement and another for background suppression then
obtain the difference. \cite{Cernadas1998DetectionMC} apply the same idea but, rather, by using a
Directional Recursive Median Filter (DRMF). \cite{Shi2018HierarchicalPipelineForBreast} use novel
texture features derived from combined Law's texture features.

\subsubsection{Pixel-based}\label{sec:pixel-based}
\cite{Chan1995MammographicMicrocalcification} is an early attempt to use CNN. Later, the authors
elaborated in \cite{Gurcan2002OptimalNeuralNetworkArchitecture} on the CNN architecture and enhanced
the accuracy. The authors elaborated more on the approach in
\cite{Ge2006ClusterMicrocalcification}. Instead of using the score image output of the final layer
of the CNN as a final classifier, they combined it with the morphological features of
microcalcification clusters to form a new set of features, which are fed to a simple LDA
classifier. They report one of the most accurate results in the literature. \cite{Elnaqa2002SVM} use
featureless approach described above (Sec.~\ref{sec:pixel-based-approach}) with the Support Vector
Machine (SVM) but on the output image of a High-Pass Filter (HPF) rather than the image itself. The
HPF is to suppress the low frequency components that cannot be a microcalcification
focus. \cite{Wei2005RVM} follow the same route but with replacing the SVM by Relevance Vector
Machine (RVM) for faster computations. Both results are similar and are two of the most accurate
results in the literature, as well, for microcalcification detection.

\subsection{Assessment}\label{sec:background-assessment}
\subsubsection{Studies, Trials, and General Issues}\label{sec:studies-overview}
Since CAD should be used as a second reader that assists radiologists (hence, the word ``Aided'' in
CAD), there are many studies and clinical trials that report how the use of CAD affects
radiologists' reading accuracy. \cite{Chan1990ImproveCAD, Chan1999Improvements,
  Jiang1999ImprovingBreast, Freer2001ScreeningMammography, Petrick2002EvaluationMass,
  Hadjiiski2004Improvements, Ko2006ProspectiveAssCAD, Gilbert2008SingleReading} asserts that CAD
improves radiologists' accuracy. However, \cite{Gur2004ChangesIn} concludes that CAD does not
improve radiologists' accuracy. On the other extreme, \cite{Fenton2007InfCADMammography} (a study
published in the \textit{New England Journal of Medicine} and introduced by
\cite{Hall2007BreastImagingCAD}) concluded that CAD reduced radiologists' accuracy!
\cite{Balleyguier2005CADmammography} shed the light on this discrepancies, where they explain that
CAD improved the accuracy of junior radiologists while it slightly improved the accuracy of seniors.

\cite{Harris1997b,Krupinski2008b}, from the Radiology community, provide a pilot view on breast
cancer screening and medical assessment of imaging system and the effect of using CAD. In addition,
they suggest the best practice for assessment strategy for any imaging system, including hardware,
software, time measurements, ROC analysis, etc.

\bigskip

The accuracy of a particular CAD is measured conditionally on a particular dataset; hence, it is the
average over lesion sizes, types, and breast densities available in this dataset. Therefore,
conditioning on a particular lesion size, type, or breast density will produce different
accuracies. \cite{Malich2003Influence, Brem2007ImpactOfBreastDensity} study the impact of breast
lesion size and breast density on the measured accuracy. \cite{Soo2005CADamorphusCalc} (a study
published in American Journal of Roentgenol (AJR) and discussed by \cite{Hall2006CADamorphousCalc}
in the same journal) studied the accuracy of one of the available commercial CADs (Image checker
3.2) to detect amorphous calcifications. They reported that the accuracy is markedly lower than
previously reported for all malignant calcifications. \cite{GarciaManso2013ConsistentPerformance}
designed their CAD so that the accuracy does not vary across different datasets. They leveraged
DDSM, which is acquired from 4 different institutions, to train on one institutional dataset and
test on the all three remaining. They asserted that the accuracy variance of this procedure is
minimal.

\subsubsection{Assessment in Terms of  FROC}\label{sec:assessm-terms-froc}
The \textit{de facto} of the field of accuracy assessment is to report the accuracy of CADs and
radiologists in terms of the conventional Receiver Operating Characteristic (ROC) or, more
elaborately and accurately, the localized version Free-response ROC
(FROC). \cite{Kallergi1999EvaluatingPerf, Baldi2000AssessingAccuracy,
  Chakraborty2004ObserverStudies, Wagner2007AssMedImgTutorial, Yoon2007Evaluating,
  He2009AlphabetSoup, Wunderlich2012NonparametricLROC} are good sources for the theory of assessment
in terms of ROC and FROC. It is quite important to clarify some terminologies and notations that are
usually used loosely in the literature. This paves the road for the assessment in
Section~\ref{sec:assessment}.

A case (or study) is a collection of mammograms taken for the same patient (usually four, two per
side). The case may be malignant/Positive (P) or normal/Negative (N), and similarly the image or the
side, depending on whether it contains a lesion marked by the radiologist and proven by biopsy or
not, respectively. If a CAD/radiologist gives a mark inside the label boundary of a positive lesion,
this detection is called a True Positive (TP). The True Positive Fraction (TPF), called as well
sensitivity, is the ratio between the number of TP detection over the number of total
positives. However, there are four levels of defining what is ``positive'': case, side, view, or
lesion; for each there is a corresponding definition of TPF as follows.
\begin{definition}[True Positive Fraction (TPF)] For each of the following criteria, the TPF is
  defined as the number of:\label{sec:assessm-terms-froc-1}

  \begin{description}[itemsep=0in]
    \item[(``per case'' or ``per study'')] cases truly detected (where a positive case is counted as
    detected if at least one lesion is detected in at least one view of at least one side of this
    case) normalized by the total number of positive cases.

    \item[(``per side'' or ``per breast'')] sides truly detected (where a positive side is counted as
    detected if at least one lesion is detected in at least one view of this side) normalized by the
    total number of positive sides.

    \item[(``per image'' or ``per view'')] images (views) truly detected (where a positive image is
    counted as detected if at least one lesion is detected in this image) normalized by the total
    number of positive images.

    \item[(``per label'')] labels truly detected (where a positive label is a radiologist single
    marking for a lesion in a single view) normalized by the total number of positive labels. This is
    a very aggressive assessment measure, where it accounts for several lesions in one view. In that
    case, a view with two lesions, where only one of them is detected will be counted as 1/2; as
    opposed to the ``per view'' assessment, where it would be counted as 1/1.
  \end{description}
\end{definition}

On the other hand, the specificity is measured in terms of the number of False Markers per Image
(FM/Image). It is measured by counting the number of total false markers produced by the CAD on
normal images normalized by the total number of normal images. A single pair of FM/Image and its
corresponding TPF constitute a single Operating Point (OP) on the FROC. Then, varying the level of
reading aggressiveness produces higher OPs up the curve.

\subsubsection{Assessment in Terms of Other Accuracy Measures}\label{sec:assessm-terms-other}
Rather than the \textit{de facto} of the field, ROC and FROC, there are other accuracy measure
indices. We will just give two of them as examples. For the assessment of region based methods,
where the task can be lesion segmentation, a common accuracy measure is the Intersection over Union
(IoU). It is defined as the intersection (in pixels) of the true lesion and detected one normalized
by their union.

For the assessment of pixel based methods, in \cite{Yousef2010OnDetecting} we suggested the use of
the MannWhitney statistic directly on the score image before postprocessing. The AUC of a score
image is defined as%
\begin{align}
  \widehat{AUC}   =  \frac{1}{n_1n_2}\sum_{i=1}^{n_1}{\sum_{j=1}^{n_2}{I_{(x_i < y_j)}}},\ I_{(c)} = \left\{\begin{array}[c]{ll} 1, & \text{if c is true} \\0, &\text{if c is false}\end{array}\right.\label{eq:AUC},
\end{align}%
where $x$ is the set of scores of the normal region, $y$ is the set of scores of the malignant
region(s), and $n_1$ and $n_2$ are the respective number of pixels. This is a measure of the
separability of the two sets of scores. The closer the AUC to a value of 1 the more the separability
between the two sets of scores, and hence the more accurate the classifier on this image.


%% file: SecAlgorithms.tex
\section{Method: Design}\label{sec:algorithms}
In this section we detail the design steps of our
CAD. Sections~\ref{sec:segmentation}--\ref{sec:micr-detect} discuss the four main sequential steps:
breast segmentation, image enhancement, mass detection, and microcalcification detection
respectively. The method assessment is deferred to Section~\ref{sec:assessment}.

To leverage the training dataset, as much as possible in building the algorithms, all values of
tuning and complexity parameters in this design process are obtained using the Leave-One-Out
Cross-Validation (LOOCV) rather than the typical 10-fold CV. This will not influence the classifier
stability since, our assessment (Figure~\ref{fig:FROCs}) measures the ``conditional'' accuracy,
i.e., conditional on the training dataset. This is the actual CAD that will be revealed to
radiologists. For elaborate discussion on the difference among the ``accuracy'', ``true accuracy'',
``conditional accuracy'', and other versions, the reader can refer to
\cite{Yousef2004ComparisonOf,Yousef2006AssessClass, Yousef2019AUCSmoothness-arxiv,
  Yousef2019PrudenceWhenAssumingNormality-arxiv}.

\subsection{Segmentation}\label{sec:segmentation}
We do not remove the pectoral muscle since we found very few cases with lesions in it. Therefore,
our segmentation algorithm starts with Otsu's method \citep{Otsu1979ThresholdSelectionMethodFrom} on
the whole mammogram, and takes its output threshold, which classifies the image to two classes
(breast and background), as an initial seed. Then, in each iteration we calculate the Quadratic
Discriminant Analysis (QDA) score for each pixel in these two classes and dismiss any pixel whose
QDA score and its comparison with its class mean agree. The iterative algorithm converges when the
number of dismissed pixels is less than a stopping criterion that is equal to 0.002 of the total
number of image pixels. The algorithm is very successful in separating the breast from the
background with no overlapping region. Figures~\ref{fig:image-enhancement}--\ref{fig:Mass-Centroids}
presents examples of segmented mammograms.

\subsection{Image Enhancement}\label{sec:image-enhancement}
Mammograms acquired from different mammography machines are of different clarity and quality for
both human visual inspection and detection algorithms. Therefore, we do image enhancement in a novel
way so that mammograms will look almost of the same qualitative clarity to radiologist and of the
same quantitative standardization to the detection algorithms regardless of the producing
mammography machine. Figure~\ref{fig:image-enhancement} is an example of the final
\texttt{EnhancedImage} returned by the image enhancement algorithm~\ref{lst:Enhancement} that will
be explained below.

\bigskip

Global Histogram Equalization (GHE) is common in some literature (reviewed in
Sec.~\ref{sec:background-image-enhancement}); however, our novel method relies on Local Histogram
Specification (LHS) as opposed to GHE or LHE. Using a sliding window, we ``locally--specify'' (or
locally--enforce) the histogram of each pixel and its neighbors, within a vicinity, to be of a
specific probability distribution with fixed chosen parameters. With this simple, yet novel, idea
mammograms regardless of the producing mammography machine will have all of their pixels with the
same local histogram distribution with the same parameters (it is clear that LHE is the same as LHS
with Uniform distribution). We experimented with several probability densities, including
Exponential, truncated Exponential, Gamma, and Uniform, we found that the exponential distribution
achieves both the best visual quality and the best detection accuracy. The insight behind the
exponential distribution is that it imposes more separation between the low gray level pixels and
high level ones. Since cancer, whether masses or microcalcification, is depicted in mammograms as
those set of pixels with higher gray level than their neighbors, the exponential distribution then
helps in emphasizing this discrepancy. Quantitatively, the detection algorithm that possessed
highest accuracy in terms of the FROC assessment (Sec.~\ref{sec:assessment}) is the exponential
distribution, which supports this argument. The choice of the exponential distribution, along with
the tuning parameters discussed below, are chosen using the LOOCV.

\bigskip

Algorithm~\ref{lst:Enhancement} presents the steps of our image enhancement; the steps are explained
as follows. As long as images come from different mammography machines, for an image $m$ we
\texttt{ScaleImage} to a standard height of 230 pixel; the width is scaled such that the aspect
ratio is preserved. We, then, \texttt{SegmentImage} to separate the breast from the background
(Sec.~\ref{sec:segmentation}). \texttt{NumPixels} is the \texttt{BreastSize}, where each of these
pixels has a gray level $0 \leq t < 255$. For the $i\textsuperscript{th}$ pixel of interest (POI),
$0 \leq i < \mathtt{NumPixels}$, take a surrounding window of size $\mathtt{W}=81$ centered around
$i$; this region will be of size $\mathtt{W}^2$ pixels and the window will be sliding to cover all
POIs in the breast region (Figure~\ref{fig:wind-surr-pixel}). We do
\texttt{LocalHistogramSpecification} of this region to set the new pixel level of the
$i\textsuperscript{th}$ POI and loop over all pixels. The PDF of the window surrounding the POI,
i.e., the histogram of 256 bins (a bin for each gray level value $t$) is denoted by
\texttt{PDFwin(t)}. Its Cumulative Distribution Function (CDF) is denoted by \texttt{CDFwin(t)}. A
discrete exponential density function with $\mathtt{lambda} = 10$, and 256 bins as well, is created
and denoted by \texttt{PDFexp(t)}; its CDF is \texttt{CDFexp(t)}. Now, we transform the PDF of the
region surrounding the $i\textsuperscript{th}$ POI from \texttt{PDFwin} to \texttt{PDFexp} by
applying the transforms of Eq.~\eqref{eq:4} to the gray
levels. 
\begin{subequations}\label{eq:4}
  \begin{align}
    tnew      & = CDFexp^{-1}\bigl( CDFwin(t) \bigr),\label{eq:1} \\
    CDFwin(t) & = \sum_{x=0}^{x=t} PDFwin(x),       \label{eq:2}   \\
    CDFexp(t) & = \sum_{x=0}^{x=t} PDFexp(x).\label{eq:3}
  \end{align}%
\end{subequations}

\subsection{Mass Detection}\label{sec:mass-detection}
The mass detection algorithm is pixel-based and featureless
(Sec.~\ref{sec:background-mass-detection}). It trains and tests
(Algorithms~\ref{lst:Training}--\ref{lst:Testing}, respectively) directly at the pixel
level. Figure~\ref{fig:mass-micro-detect} illustrates the two outputs of these Algorithms. The
figure illustrates the \texttt{ScoreImage}, where each pixel is assigned a score value that
corresponds to its likelihood of being malignant. Therefore, high scores (bright regions) correspond
to cancerous regions and low scores (black regions) correspond to normal ones. The figure
illustrates as well the \texttt{MassLocation}, which is a red marker(s) plotted on the original
image to indicate the center of a cancer.

\bigskip

Algorithm~\ref{lst:Training} presents the steps of the training phase. All tuning and complexity
parameters discussed below are obtained using LOOCV. Each image \texttt{m}, where
$0 \leq \mathtt{m} < \mathtt{M}$, is scaled down, segmented, then enhanced using
Algorithm~\ref{lst:Enhancement}. For the $i\textsuperscript{th}$ POI,
$0 \leq i < \mathtt{NumPixels}$, we take a surrounding window of size $\mathtt{w_1}=21$ (less than
the $\mathtt{W}=81$ used for enhancement) centered around $i$. This region will be of size $w^2=441$
pixels and the window will be sliding to cover all POIs in the breast region
(Figure~\ref{fig:wind-surr-pixel}). For each POI $i$, its features will be all the 441 gray levels
of the surrounding window. For each training image \texttt{m}, the breast is divided into two
regions: mass (with number of pixels $\mathtt{NumPixlesMass}$) and normal (with number of pixels
$\mathtt{NumPixelsNormal}$), where of course
$\mathtt{NumPixels} = \mathtt{NumPixelsMass} + \mathtt{NumPixlesNormal}$. For each image a
\texttt{MassFeatures} matrix of size $\mathtt{NumPixelsMass} \times 441$ is constructed. Each row in
this matrix represents the 441 gray levels of the surrounding window of the corresponding POI. Then
a $K$-means clustering is used to obtain the best 100 representatives to reduce the number of rows;
i.e., the size of the matrix will be reduced to $100 \times 441$. For all training images
\texttt{M}, the $100 \times 441$ mass centroid matrix of each of them are stacked together to
construct one larger matrix \texttt{MassCentroidList} of total size of $(100\mathtt{M} \times
441)$. Similarly, the matrix \texttt{NormalCentroidList} is constructed from the normal regions of
the same \texttt{M} images. As we reduced the number of rows to find the best representatives, we
reduce the number of columns (the 441 features) to find the best features. We obtain this by running
Principal Component Analysis (PCA) on the $(200\mathtt{M} \times 441)$ joined matrix
[\texttt{MassCentroidList}; \texttt{NormalCentroidList}], and take the largest 10 components
($\mathtt{C}=10$) to be the new reduced set of features (or \texttt{PCVectors}). The two matrices
now are projected to the new vector space and, hence, reduced in size to $100\mathtt{M} \times 10$
each. These two matrices will be used to give a score to each pixel in a future (testing) image
(Algorithm~\ref{lst:Testing}).

\bigskip

Algorithm~\ref{lst:Testing} presents the steps of the testing phase. Given a testing image, the
first step is to \texttt{EnhanceImage} (Algorithm~\ref{lst:Enhancement}). Then, for each pixel in
the breast we get its 441 feature vector; this is once again the plain gray level of the pixels in a
surrounding window of size $21 \times 21$. This 441 feature vector is projected on the 10
\texttt{PCVectors}, obtained from the training phase, to produce only 10 features. Then, each pixel
is compared to the two extracted matrices from the training phase, \texttt{MassCentroidList} and
\texttt{NormalCentroidList}, using a KNN algorithm with chosen $\mathtt{K}=141$. The KNN algorithm
gives a score to each testing pixel instead of a binary decision; the score is $\log(n_1/n_2)$,
where $n_1$ and $n_2$ are the number of closest training observations in each class. The global set
of scores are normalized to values between 0 and 1 to be used as a probability of malignancy. After
running the algorithm on all breast pixels of the testing image, this produces a
\texttt{ScoreImage}, which will be finally post-processed to obtain the center of the cancerous
mass. The post processing is the \texttt{SmoothImage} step. It is a mere $10 \times 10$ smoothing
filter that runs over the breast region of the \texttt{ScoreImage}. After smoothing, we search the
smoothed image to \texttt{FindMaxima}. The location of these maxima are the \texttt{MassLocation}
finally returned by the algorithm. The scores of these locations, \texttt{ScoreImage(MassLocation)},
are the probability of having cancer at these locations.

\bigskip

Finally, when using the CAD, the radiologist can see the score image beside, or overlaying, the
original image, in addition to the red marker on the cancer location
(Figure~\ref{fig:mass-micro-detect}). Radiologists can set a threshold (level of aggressiveness) to
trade-off sensitivity (TPF) vs FM/Image; all centers of masses with scores lower than this level
will be ignored.

\subsubsection{Tuning and Complexity Parameters}\label{sec:tuning-compl-param}
\begin{figure}[!tb]
  \centering
\if 0\MYJOURNAL%
  \includegraphics[width=0.35\textwidth, trim={0 0 0 30},clip]{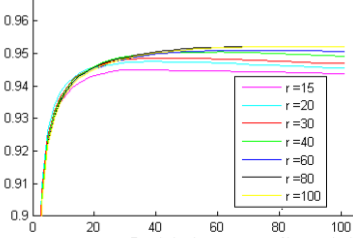}\hfil\includegraphics[width=0.35\textwidth]{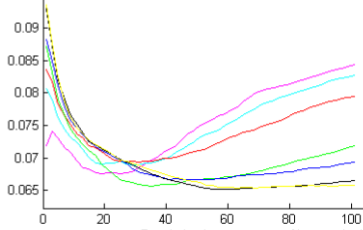}
\else \if 1\MYJOURNAL%
  \includegraphics[width=0.25\textwidth, trim={0 0 0 30},clip]{./Graphics/KNN-AUC-Mean-Final.png}\hfil\includegraphics[width=0.23\textwidth]{./Graphics/KNN-AUC-Std-Final.png}
\fi\fi
\caption{The plain effect---without preprocessing (enhancement) or postprocessing (score image
  smoothing)---of the number of centroids per image $r$ and KNN nearest neighbors $K$ on the
  separability of the normal and abnormal pixels measured in AUC (left) and its standard deviation
  (right). For each $r$, there is $K$ that is optimal in terms of the maximum AUC w.r.t. its
  standard deviation. The optimal $K$ increases with $r$.}\label{fig:plain-effect-without}
\end{figure}
As introduced above, the LOOCV was used to chose all tuning and complexity parameters just explained
above: the image enhancement window size ($\mathtt{W} = 81$), the exponential distribution parameter
($\mathtt{lambda} = 10$), the mass detection window size ($\mathtt{w_1} = 21$), the number of
centroids per image ($\mathtt{r} = 100$), the number of largest PCs ($\mathtt{C} = 10$), and the KNN
parameter ($\mathtt{K} = 141$). For the sake of demonstration, Figure~\ref{fig:plain-effect-without}
illustrates the ``plain effect'' of the number of $K$-means clustering centroids \texttt{r}, which
is a tuning parameter, and the number of nearest neighbors \texttt{K}, which is a complexity
parameter, on the classifier accuracy measured in terms of the AUC (Eq. \eqref{eq:AUC}) and its
standard deviation. ``Plain effect'' means that neither preprocessing (enhancement) nor
postprocessing (score image smoothing) is involved. It is clear that for a particular \texttt{r},
there is an optimal value of \texttt{K} that achieves a maximum AUC, a common phenomenon for any
classifier complexity parameter. In addition, the optimal choice of \texttt{K} increases with
\texttt{r}, which is expected as well since increasing \texttt{r} increases the final size of the
data matrix. The same phenomenon happens with increasing the dataset size \texttt{M} for the same
reason.

\subsubsection{Multi Classifier System (MCS)}\label{sec:multi-class-syst}
It is reasonable to anticipate that varying the feature window size $\mathtt{w_1}$ would be able to
detected other masses of different size; this is a sort of detection at multi-resolution levels. An
MCS version of Algorithms~\ref{lst:Training}--\ref{lst:Testing} is constructed by simply running it
for 10 different window sizes $\mathtt{w_1} = 6m + 3,\ m=1 \cdots 6$, where $\mathtt{w_1} = 21$ is
one among them. For each of these 10 windows, a score image is produced for a single mammogram; then
the final score image is the average of all of these score images. The smoothing step of the final
averaged score image is done as above. The MCS version increases the detection accuracy remarkably
(Figure~\ref{fig:FROCs}) as will be discussed in Sec.~\ref{sec:assessment}.

\subsubsection{Other Classification Methods and Features}\label{sec:other-class-meth}
The KNN classifier was the overall winner in hundreds of experiments in which we varied classifiers
and their complexity parameters. These classifiers are NN with one-layer, two-layers, and different
number of neurons; SVM with different number of kernels; different versions of KNN, e.g., DANN
\citep{Fukunaga1984OptimaGlobalNN, Short1981OptimalMeasureNN}; Classification and Regression Trees
(CART) with varying number of parents and pruning levels; and Random Forest (RF) with varying number
of trees. For each classifier, a similar behavior to that of Figure~\ref{fig:plain-effect-without},
is obtained. The overall winner was the KNN explained above.

In addition to various classifiers, we experimented various pixel-based methods
(Sec.~\ref{sec:pixel-based-approach}) rather than our feature-less pixel-based approach. We have
reported in \cite{Yousef2010OnDetecting} the accuracy of the KNN classifier measured in AUC for some
of these approaches including IRIS filter and template matching. Our feature-less pixel-based
approach along with the KNN classifier is the overall winner among all other features and
classifiers.

\subsection{Microcalcification Detection}\label{sec:micr-detect}
The microcalcification detection algorithm~\ref{lst:Microcalcification} detects and marks individual
microcalciﬁcation foci, even if they are not clustered. Figure~\ref{fig:mass-micro-detect} is
illustrates the output \texttt{FociLocation} of the microcalcification detection algorithm.

One of the well known accurate results in the literature for detecting microcalcifications, and
discussed in Sec.~\ref{sec:pixel-based}, is~\cite{Elnaqa2002SVM}. After investigating their
algorithm and reproducing their results we concluded that the naive preprocessing step, prior to the
SVM training and testing, is almost what is responsible for the high accuracy---not the
classification phase using the SVM. This drove our motivation to design a more sophisticated image
processing step, a two-nested spatial filter, to detect the microcalcifications. So, our
microcalcification detection algorithm, as opposed to the literature and as opposed to our mass
detection algorithm, is based solely on image processing with no machine learning training and
testing phases.

\bigskip

Algorithm~\ref{lst:Microcalcification} presents the steps of our algorithm of microcalcification
detection. The idea of the algorithm is to detect those pixels whose very few close neighbors have
high gray level, and the next surrounding neighbors have much lower level. This is essentially what
characterizes the microcalcification foci. Since the size of a microcalcification focus may reach
1/1000 of the mammogram size, we scale all mammograms to a standard height of 2294 pixel, as opposed
to the small scaling of 230 pixels of mass detection. The microcalcification detection algorithm
scans the breast region, pixel by pixel, with sliding filter composed of two nested filters:
\texttt{InnerFilter} (with positive coefficients and size $\mathtt{w_2} \times \mathtt{w_2}$) and
\texttt{OuterFilter} (with negative coefficients and size $3\mathtt{w_2} \times 3\mathtt{w_2}$) as
indicated in Figure~\ref{fig:wind-surr-pixel}. This filter detects the gray level contrast between
high gray level POI (inner) and its low gray level surrounding neighbors (outer). The difference
between these two nested filters is a measure of this contrast at each POI. In practice instead of
looping on the image pixel by pixel to apply the filter we do it directly in the frequency domain
using Fourier Transform for faster execution. When the contrast exceeds a particular threshold
\texttt{Th}, the POI is a candidate microcalcification focus. As indicated in
Figure~\ref{fig:wind-surr-pixel}, the boundary distance between the two filters from each side is
$\mathtt{w_2}$, which equals to the size of \texttt{InnerFilter} itself. This ensures comparing the
bright pixels of particular width with their surrounding neighbors of the same width. Typical chosen
values, after many experimentation, are $\mathtt{w_2} = 2$ and $\mathtt{Th} = 2.4$. After applying
the threshold \texttt{Th} we do 8-connected region to join 8-connected pixels to just a single one
at their center, which will be the final mark of a single detected microcalcification focus.

A final remark on the values of the inner and outer filter (Figure~\ref{fig:wind-surr-pixel}) is
due. There is only one constraint on the filter design, i.e., the summation across all of the filter
pixels is zero. Since the inner filter is of size $\mathtt{w_2}$ and the outer filter is of size
$3\mathtt{w_2}$, they enclose $\mathtt{w_2}^2$ and $8\mathtt{w_2}^2$ pixels respectively. Hence, the
condition is $a \mathtt{w_2}^2 + 8 b \mathtt{w_2}^2 = 0$, which is equivalent to $a/b = -8$. The
chosen values of $1/4$ and $-1/32$, indicated in the Figure, are therefore just one possible candidate.

\bigskip

Finally, the following procedure is defined to cluster the detected foci. This is necessary for
defining a threshold (level of aggressiveness) for the trade off between sensitivity and FM/Image,
for objective assessment. The procedure starts by assigning a cluster to each individual detected
focus. Then two clusters are merged together if there are two foci, one from each cluster, that are
less than 3mm apart. The procedure repeats until convergence when no more cluster can be
merged. Now, each cluster has a number of foci with a geometric center. The center is the final mark
that is displayed as a positive finding if the number of foci in the cluster is higher than the
selected threshold. If the marker gets inside the boundary of a true microcalcification cluster
marked by the radiologist and proven by biopsy it is considered as a TP otherwise it is a FM.

\subsection{GPU Implementation and Computational Acceleration for Algorithms and
  Experiments}\label{sec:gpu-comp-accel}
In a project of this size, where many algorithms were tested and thousands of experiments were
conducted we had to reduce the very expensive run time to acceptable bounds from a practical point
of view. Three different levels of difficulties are encountered in this regard.

First, the majority of the code written for this project was in Matlab; and some codes were written
in C++ for computation speedup (we reported the performance comparison between Matlab and OpenCV in
\cite{Elsayed2019Matlab-arxiv} using 20 different datasets than the dataset of our present
project). At the time of this work, MATLAB did not have the currently available OpenCL
toolbox. Therefore, we wrote an opensource wrapper for MATLAB, which automated the process of
compiling OpenCL kernels, converting MATLAB matrices to GPU buffers, converting data formats and,
reshaping the data. This wrapper made it much easier to run, experiment, and deploy our custom
kernels with a few lines of MATLAB code, and zero lines of C++ code. This wrapper is available as an
opensource at~\cite{LIBCADUtil2012}.

Second, some algorithms were, tediously, very computationally intensive, e.g., the LHS and the KNN
(\ref{sec:image-enhancement}) since they iterate over all pixels in the dataset. Therefore, we
developed GPU versions of these algorithms to take full advantage of the GPU architecture and to
leverage different levels of memory and caching. The algorithm had a significant speedup of up to
120X


Third, the number of total configurations and experiments were in thousands, since it is the cross
product of image enhancement parameters, feature parameters, classifier complexity parameters,
etc. Therefore, We developed an in-house cluster and experiment management system; it is available,
as an opensource, at~\cite{LIBCADUtil2012}. At the time of this work, big data processing frameworks
such as Spark and Hadoop were still infant, and did not support MATLAB. Our in-house system allowed
us to operate a Beowulf cluster, run numerous experiments and have their configurations and results
tracked and stored in an efficient and effective manner. Our system was akin to Sacred
\citep{Sacred2017}; only it also provided provisioning and resource allocation.

%% file: SecAssessment.tex
\section{Method: Assessment}\label{sec:assessment}
This section reports the assessment of the mass and microcalcification detection algorithms
(Sec.~\ref{sec:algorithms}). There are three different versions of the algorithms: without LHS
enhancement, with LHS enhancement, and the MCS version. For easy referencing, these three versions
are denoted by \MYCAD~1, 2, and 3 respectively---the name~\MYCAD~is explained in
Sec.~\ref{sec:deploying-cad-as}.

On the other hand, to compare this CAD to other commercial CADs, the optimal practice is to test
them all on the same dataset; this is, obviously, to have common difficulty and heterogeneity of
images to avoid finite dataset size performance variance. However, comparing \MYCAD~to the
commercial CADs is almost impossible since it requires purchasing a license for each. Moreover, some
of these commercial CADs are not deployed except with the corresponding vendor's mammography
machine. The U.S. FDA had approved the use of CADs in 1998; since then many CAD systems have been
developed and approved. The literature has few published clinical trials that report the accuracy of
some of these CADs. However, many of these studies report a single operating point on the FROC (a
pair of sensitivity and false markers per image) of one version of a particular CAD. Moreover, not
all studies report the accuracy of both mass and microcalcification detection. To the best of our
knowledge, there is no comprehensive comparative study for all of these CADs. It is then very
informative, even for its own sake in addition to the comparison to \MYCAD, to compile the results
of all of these studies and summarize them on a single FROC; then present them along with the
assessment results of \MYCAD~for comparison.

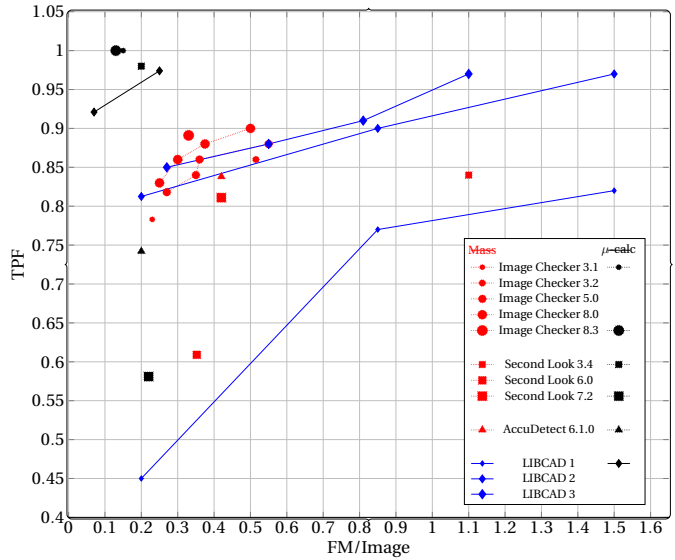
\begin{figure}[!tb]
  \centering
  \input{CommercialCAD.tex}
  \caption{FROCs of mass and microcalcification detection of three versions of \MYCAD, as tested on
    the project sequestered validation set, in comparison with the most recent U.S. FDA approved and
    commercial CADs (5 versions of Image Checker, 3 versions of Second Look, and AccuDetect), as
    appears in literature and corresponding official documents---numbers are compiled and detailed
    in Tables~\ref{tab:comm-cads-results}--\ref{tab:comm-cads-comments}. \textbf{Mass:} \MYCAD~1
    (and 2) are without (and with) the image enhancement via Local Histogram Specification
    (LHS). The effect of LHS on increasing the TPF at each FM/Image is amazing. \MYCAD~3 is the MCS
    version of \MYCAD~2. The effect of MCS is remarkable as well. \MYCAD~3 is slightly outperformed
    only by Image Checker 8 and versions above. \textbf{Microcalcification:} the three versions of
    \MYCAD~have almost the same accuracy, and hence only that of \MYCAD~2 is illustrated. The figure
    illustrates that its sensitivity is very close to three versions of commercial CADs and highly
    outperforming two other versions.}\label{fig:FROCs}
\end{figure}

Sec.~\ref{sec:different-versions} reports the assessment results of \MYCAD,
Sec.~\ref{sec:comp-with-comm} compiles the assessment results of these commercial CADs, and
Figure~\ref{fig:FROCs} presents all of these results comparatively.

\subsection{\MYCAD}\label{sec:different-versions}

\subsubsection{Assessment of Mass Detection}\label{sec:assessm-mass-detect}
All results are reported using a testing dataset that has never been part of the training process to
avoid any overtraining. 
The FM/Image are obtained from testing on the normal images, whereas the sensitivity (TPF) is
obtained from testing on the positive cases. Each operating point on the FROC is calculated by
setting a threshold for the score of points of maxima of the final \texttt{ScoreImage}
(Sec.~\ref{sec:mass-detection}). At a particular threshold, an operating point of (FM/Image, TPF) is
calculated as explained in Sec.~\ref{sec:assessm-terms-froc}.

Figure~\ref{fig:FROCs} illustrates that the mass detection algorithm of \MYCAD~1 has the worst
accuracy among others. The FROC of \MYCAD~2 is remarkably higher at all FM/Image values, and in
particular at the lower range where the difference is almost 35\% TPF at 0.2 FM/Image. The effect of
the LHS enhancement preprocessing step is remarkable in boosting the sensitivity. \MYCAD~3 is the
MCS version with LHS enhancement to enable detecting masses of different sizes. Compared to
\MYCAD~2, there is a clear improvement of sensitivity at all FM/Image values. When comparing the
three versions of \MYCAD~to the commercial and FDA approved CADs (Sec.~\ref{sec:comp-with-comm}), it
is clear that \MYCAD~1 is the worst, \MYCAD~2 has a midway sensitivity performance, whereas \MYCAD~3
achieves the highest sensitivity if compared to all others except Image Checker (ver. 8 and
ver. 8.3, but not earlier versions) only at small range of FM/Image. However, at values larger than
0.5 FM/Image, no sensitivity is reported for Image Checker.

Figure~\ref{fig:Mass-Centroids} presents examples of true and false positives. For all displayed
views, we kept decreasing the threshold until a marker appeared inside the lesion. Markers lie
outside lesions are FP. The first three views are from three different cases. The last two views are
from the same case. The lesion in this case is detected in one view from the first marker and not
detected up to the second marker in the other view.

\bigskip

The TPF reported for the three versions of LIBCAD (1, 2, and 3) in Figure~\ref{fig:FROCs} is ``per
study'' and only using database D1 that is explained in
Table~\ref{tab:datab-acqu-design}. Table~\ref{tab:OP-DataBases} takes LIBCAD2 as an example to
illustrate the behavior on different databases (D1, D2, and D3) and using different TPF assessment
strategies (``per case'', ``per side'', and ``per image''). It is remarkable that at the mid and
upper range of the FROC (FM/image of 0.85 and 1.5), the TPF is almost constant across databases,
which asserts the effect of the LHS enhancement method that standardizes the probability
distribution of the images. Yet, at the lower scale of the FROC (FM/image of 0.2) there is a TPF
difference up to only 10\%.

\begin{table}[!tb]
  \centering
  \begin{tabular}{llll|lll|lll}
    \toprule
    & \multicolumn{3}{c}{\textbf{D1}} & \multicolumn{3}{c}{\textbf{D2}} & \multicolumn{3}{c}{\textbf{D3}}\\
    \midrule
    \textbf{FM/image}     & 0.2     & 0.85    & 1.5  & 0.2    & 0.85    & 1.5  & 0.2    & 0.85   & 1.5     \\
    \midrule
    \textbf{TP Per Case}  & 81.25 & 90    & 97 & 78   & 91.4  & 94 & 73   & 89   & 90    \\
    \textbf{TP Per Side}  & 82.4  & 91.2  & 97 & 76.2 & 90.5  & 93 & 71.5 & 87.5 & 91.2  \\
    \textbf{TP Per Image} & 70    & 91.25 & 95 & 66.3 & 87.25 & 90 & 63.3 & 81.7 & 86.75 \\
    \bottomrule
  \end{tabular}
  \caption{Operating points using different TPF measures (Definition~\ref{sec:assessm-terms-froc-1})
    and using different databases (Table~\ref{tab:datab-acqu-design}). The CAD sensitivity, at each
    FM/image, is consistent among databases.}\label{tab:OP-DataBases}
\end{table}

\subsubsection{Assessment of Microcalcification Detection}\label{sec:assessm-micr-detect}
Although the dataset of this project contains more than 1500 cases, the number of cases with
microcalcifications is only 38 cases (67 images). For this reason, along with the fact that the
microcalcification detection algorithm is not a machine learning based on training-and-testing
paradigm, all of these positive cases are used to estimate the sensitivity. The FM/Image is measured
on the normal cases exactly as was done for the mass detection assessment. \MYCAD~detects and marks
microcalciﬁcation foci, even if those foci that are not clustered (Sec.~\ref{sec:micr-detect}). We
published a detailed study of the accuracy of \MYCAD\ 1 compared to the expert radiologist in
\cite{AbdelRazek2013MicroclacificationLIBCAD}. The study was addressing the Radiology community and
did not include technical or algorithms details. Figure~\ref{fig:FROCs} illustrates that at 0.25
FM/Image, the sensitivity is 0.974 (=37/38; i.e., it misses just one image out of 38). At 0.07
FM/Image, the sensitivity is 0.921 (=35/38; i.e., it misses two more images). Since the missed cases
for \MYCAD~1 are very few, the accuracy improvement of \MYCAD~2, with LHS, is marginal and not
significant and hence omitted. From these results, it seems that microcalcification detection is
much easier in detection than mass detection except in subtle cases where a microcalcification
cluster is obscured by a mass in the same case. The case missed by \MYCAD~1, as reported in
\cite{{AbdelRazek2013MicroclacificationLIBCAD}} is of this type. To study whether the LHS of
\MYCAD~2 helps in detecting microcalcifications within masses, a large number of positive cases
having such a combination has to be collected. Figure~\ref{fig:mass-micro-detect} illustrates an
example of an image with subtle microcalcifications; however, was detected and marked by the CAD.

\subsection{Commercial and FDA-Approved CADs}\label{sec:comp-with-comm}
\begin{table}[!tb]
  \centering
  \begin{tabular}{lrcll}
    \toprule
    \textbf{ver.} & \textbf{citation}                      & \textbf{mass/$\mu$} & \textbf{TPF} & \textbf{FM/Image} \\
    \midrule
    \multicolumn{5}{c}{Image Checker, 1998}                                                                         \\
    \midrule
    3.1           & \citep{Kim2010ComparisonCommercialCAD} & mass                & 0.783        & 0.23              \\
    3.1           & \citep{Kim2010ComparisonCommercialCAD} & $\mu$               & 1            & 0.15              \\
    3.2           & \citep{Roehrig2005ManufacturerePresp}  & mass                & 0.86         & 2.06/4            \\
    5.0A          & \citep{Ellis2007EvaluationCAD}         & mass                & 0.818        & 1.08/4            \\
    5.0           & \citep{Roehrig2005ManufacturerePresp}  & mass                & 0.84         & 1.4/4             \\
    5.0           & \citep{Roehrig2005ManufacturerePresp}  & mass                & 0.86         & 1.44/4            \\
    5.0           & \citep{Roehrig2005ManufacturerePresp}  & mass                & 0.88         & 2.2/4             \\
    8.0           & \citep{Roehrig2005ManufacturerePresp}  & mass                & 0.83         & 1.0/4             \\
    8.0           & \citep{Roehrig2005ManufacturerePresp}  & mass                & 0.86         & 1.2/4             \\
    8.0           & \citep{Roehrig2005ManufacturerePresp}  & mass                & 0.88         & 1.5/4             \\
    8.0           & \citep{Roehrig2005ManufacturerePresp}  & mass                & 0.90         & 2.0/4             \\
    8.0           & \citep{Kim2010ComparisonCommercialCAD} & mass                & 0.89         & 0.33              \\
    8.0           & \citep{Kim2010ComparisonCommercialCAD} & $\mu$               & 1            & 0.13              \\
    \midrule
    \multicolumn{5}{c}{Second Look, 2002}                                                                           \\
    \midrule
    3.4           & \citep{Brem2005CAD}                    & mass                & 0.84         & 1.1               \\
    3.4           & \citep{Brem2005CAD}                    & $\mu$               & 0.98         & 0.2               \\
    6.0           & \citep{Ellis2007EvaluationCAD}         & mass                & 0.609        & 1.41/4            \\
    7.2           & \citep{Lobbes2013MalignantLesions}     & mass                & 0.811        & 1.68/4            \\
    7.2           & \citep{Lobbes2013MalignantLesions}     & $\mu$               & 0.581        & 0.88/4            \\
    \midrule
    \multicolumn{5}{c}{AccuDetect, 2013}                                                                            \\
    \midrule
    4.0.1         & \citep{Lobbes2013MalignantLesions}     & mass                & 0.838        & 1.68/4            \\
    4.0.1         & \citep{Lobbes2013MalignantLesions}     & $\mu$               & 0.742        & 0.8/4             \\
    \midrule
    \multicolumn{5}{c}{Kodak Mammography, 2004}                                                                     \\
    \midrule
    ---           & \citep{Kodak2004}                       & mass                & 0.87         & 1.0               \\
    ---           & \citep{Kodak2004}                       & $\mu$               & 0.97         & 1.0               \\
    \midrule
    \multicolumn{5}{c}{Image Intelligence, --- }                                                                      \\
    \midrule
    MV-SR6577EG   & \citep{Kuroki2012PerformanceEval}      & mass                & 0.83         & 2.5/4             \\
    MV-SR6577EG   & \citep{Kuroki2012PerformanceEval}      & $\mu$               & 1            & 0.19              \\
    \bottomrule
  \end{tabular}
  \caption{Some operating points (FROC sensitivity vs. false markers per image) of 5 commercial
    and/or FDA-approved CADs as compiled and summarized from literature. The year following the CAD
    name, e.g., ``Image Checker, 1998'', is the FDA approval year as appears on the FDA official
    site. Each row shows: a particular CAD version (ver.), the clinical trial of this CAD version
    (citation), type cancer detected (mass/$\mu$), sensitivity (TPF), and False Markers per Image
    (FM/Image). Table~\ref{tab:comm-cads-comments} includes comments and more details on these
    clinical trials. These operating points are plotted on
    Figure~\ref{fig:FROCs}.}\label{tab:comm-cads-results}
\end{table}

\begin{table}[!tb]
  \centering
  \begin{threeparttable}
    \begin{tabular}{lrrrrrr}
      \toprule
      \textbf{Citation}                                & \textbf{\#cases} & \textbf{\#masses} & \textbf{\#$\mu$-calc} & \textbf{\#mixed} & \textbf{\#normal} \\
      \midrule
      \cite{Kim2010ComparisonCommercialCAD} \tnotex{a} & 130               & 46                 & 84                     & ---                 & 130               \\
      \cite{Roehrig2005ManufacturerePresp} \tnotex{b}  & ---                 & ---                  & ---                      & ---                 & ---                 \\
      \cite{Ellis2007EvaluationCAD}                    & 192               & 192                & 0                      & 0                 & 51                \\
      \cite{Brem2005CAD}                               & 201               & 122                & 54                     & 25                & 155               \\
      \cite{Lobbes2013MalignantLesions}                & 117               & 85                 & 6                      & 26                & 209               \\
      \cite{Kodak2004} \tnotex{c}                      & 394               & 262                & 172                    & 40                & 194               \\
      \cite{Kuroki2012PerformanceEval} \tnotex{d}      & 208               & 47                 & 38                     & ---                 & 123               \\
      \bottomrule
    \end{tabular}
    \begin{tablenotes}\footnotesize
      \item[1]\label{a} The 130 normal cases are the same as the 46 + 84 masses and $\mu$-calc
      cases; however the normal breast is used. \item[2]\label{b} It is not clear how many cases are
      used in this trial. \item[3]\label{c} seems to be overall accuracy. \item[4]\label{d} it seems
      that the in-text reported accuracy and the FROC reported in that article do not agree.
    \end{tablenotes}
  \end{threeparttable}
  \caption{Information and important comments on the clinical trials that reported the accuracy of
    the commercial and FDA-approved CADs (summarized in Table~\ref{tab:comm-cads-results}). Each row
    shows: the clinical trial (citation) and number of all cases: cases with only masses, cases with
    only microcalcifications, cases with both masses and microcalcifications, and normal cases
    (respectively). The footnotes include important comments on these trials and on their reported
    accuracies.}\label{tab:comm-cads-comments}
\end{table}
The assessment results of five of the commercial CADs are compiled from several studies and
summarized in Tables~\ref{tab:comm-cads-results}--\ref{tab:comm-cads-comments}, then visualized in
the FROC of Figure~\ref{fig:FROCs}. Four of these CADs are FDA approved (Image Checker, 1998; Second
Look, 2002; AccuDetect, 2013; and Kodak Mammography, 2004), and the fifth is Image
Intelligence. Some of these CADs are included in more than one study. Also, some different versions
of the same CAD may be included in different studies. Table~\ref{tab:comm-cads-results} summarizes
this information along with the reported accuracies. The columns of this table present: the
particular CAD version (ver.), the reporting clinical study (citation), the abnormality detected
(mass/$\mu$), the sensitivity (TPF), and the corresponding specificity (FM/Image). Some of the CAD
versions have one operating point appeared in one study and others have several operating points
appeared collectively in different studies, hence constituting an FROC
(Figure~\ref{fig:FROCs}). Some TPF and FM/Image numbers in this table are not stated explicitly in
their corresponding studies; hence we calculated them and presented them explicitly in the
table. E.g., some studies reported the absolute number of detected and missed cases and absolute
number of false markers (not ratios or fractions); others reported false markers per case (not
FM/Image) and hence we divide by 4 (the number of views per case). The TPF in this table is ``per
case'' (Sec.~\ref{sec:background-assessment}). All of these studies consider cases
with only one lesion, and abandon other cases to ease calculations. This means that each case has a
single lesion in one breast (side). Therefore, these ``per case'' results are the same as the ``per
side'' ones.

Table~\ref{tab:comm-cads-comments} provides more details to the clinical studies cited in
Table~\ref{tab:comm-cads-results}. The columns of this table present: the study (citation), total
number of positive and normal cases (\# cases), number of cases with masses (\# masses), number of
cases with microcalcifications (\# $\mu$-calc), number of cases with both abnormalities (\# mixed),
and number of normal cases (\# normal). Some important comments on these studies, concerning their
results and calculation, are placed at the table footnotes.


%% file: CommercialCAD.tex
\begin{tikzpicture}[spy using outlines={circle, magnification=6, connect spies}, scale=.49]
  \pgfplotsset{every axis legend/.append style={ at={(0.81,0.025)}, anchor=south}, tick label style={font=\Large}, label style={font=\Large}}

  \begin{axis}[xlabel=FM/Image, ylabel=TPF, ymin=0.4, ymax=1.05, xmin=0, grid=major,
    width=\textwidth, legend columns=2]

    \addlegendimage{red, mark=text, text mark= Mass} \addlegendentry{} \addlegendimage{black,
      mark=text, text mark= $\mu$-calc} \addlegendentry[text depth=10pt]{}

    \addplot[red, mark=*, style=densely dotted, mark size=2pt] plot coordinates { (0.23, 0.783) };
    \addlegendentry{Image Checker 3.1} \addplot[black, mark=*, style=densely dotted, mark size=2pt]
    plot coordinates { (0.15, 1) }; \addlegendentry{}

    \addplot[red, mark=*, style=densely dotted, mark size=2.5pt] plot coordinates { (0.515, 0.86)
    }; \addlegendentry{Image Checker 3.2} \addlegendimage{empty legend} \addlegendentry{}

    \addplot[red, mark=*, style=densely dotted, mark size=3pt] plot coordinates { (1.08/4, 0.818)
      (1.4/4, 0.84) (1.44/4, 0.86) (2.2/4,
      0.88) 
    }; \addlegendentry{Image Checker 5.0} \addlegendimage{empty legend} \addlegendentry{}

    \addplot[red, mark=*, style=densely dotted, mark size=3.5pt] plot coordinates { (1/4, 0.83)
      (1.2/4, 0.86) (1.5/4, 0.88) (0.5, 0.9) }; \addlegendentry{Image Checker 8.0}
    \addlegendimage{empty legend} \addlegendentry{}

    \addplot[red, mark=*, style=densely dotted, mark size=4pt] plot coordinates { (0.33, 0.891) };
    \addlegendentry{Image Checker 8.3} \addplot[black, mark=*, style=densely dotted, mark size=4pt]
    plot coordinates { (0.13, 1) }; \addlegendentry{}

    \addlegendimage{empty legend} \addlegendentry{} \addlegendimage{empty
      legend}\addlegendentry[text depth=10pt]{}

    \addplot[red, mark=square*, style=densely dotted, mark size=2.5pt] plot coordinates { (1.1,
      0.84) 
    }; \addlegendentry{Second Look 3.4} \addplot[black, mark=square*, style=densely dotted, mark
    size=2.5pt] plot coordinates { (0.2, 0.98) }; \addlegendentry{}

    \addplot[red, mark=square*, style=densely dotted, mark size=3pt] plot coordinates { (1.41/4,
      0.609) }; \addlegendentry{Second Look 6.0} \addlegendimage{empty legend} \addlegendentry{}

    \addplot[red, mark=square*, style=densely dotted, mark size=3.5pt] plot coordinates { (1.68/4,
      0.811) }; \addlegendentry{Second Look 7.2} \addplot[black, mark=square*, style=densely
    dotted, mark size=3.5pt] plot coordinates { (0.88/4, 0.581) }; \addlegendentry{}
    \addlegendimage{empty legend} \addlegendentry{} \addlegendimage{empty legend}
    \addlegendentry[text depth=10pt]{}

    \addplot[red, mark=triangle*, style=densely dotted, mark size=3.5pt] plot coordinates {
      (1.68/4, 0.838) }; \addlegendentry{AccuDetect 6.1.0} \addplot[black, mark=triangle*,
    style=densely dotted, mark size=3.5pt] plot coordinates { (0.8/4, 0.742) }; \addlegendentry{}
    \addlegendimage{empty legend} \addlegendentry{} \addlegendimage{empty legend}
    \addlegendentry[text depth=10pt]{}


    \addplot[blue, mark=diamond*, style=solid, mark size=2pt] plot coordinates { (0.2, 0.45) (0.85,
      0.77) (1.5, 0.82)}; \addlegendentry{\MYCAD~1} \addplot[black, mark=diamond*, style=solid, mark
    size=3pt] plot coordinates { (0.25, 0.974) (0.07, 0.921)}; \addlegendentry{}



    \addplot[blue, mark=diamond*, style=solid, mark size=3pt] plot coordinates { (0.20, 0.8125)
      (0.85, 0.90) (1.5, 0.97) }; \addlegendentry{\MYCAD~2}\addlegendimage{empty legend}
    \addlegendentry{}

    \addplot[blue, mark=diamond*, style=solid, mark size=3.5pt] plot coordinates { (0.27, 0.85)
      (0.55, 0.88) (0.81, 0.91) (1.1, 0.97) }; \addlegendentry{\MYCAD~3} \addlegendimage{empty
      legend} \addlegendentry{}

  \end{axis}
\end{tikzpicture}


%% file: SecSystem.tex
\section{System: Design and Assessment}\label{sec:deploying-cad-as}
As was introduced in Sec.~\ref{sec:introduction}, commercial CADs are machine dependent; i.e.,
deployed and connected to a particular mammography machine. On the other hand, literature is full of
algorithms with no practical procedure of deploying them for public use. In this section, we
propose, implement, and assess a system architecture for deploying \MYCAD~to cloud. This system
architecture, along with the method of image enhancement (LHS of Sec.~\ref{sec:image-enhancement})
that makes \MYCAD~suitable for mammograms from any mammography machine, may participate in spreading
CAD technology.

\subsection{System Design}\label{sec:syst-arch-techn}
\begin{figure}[tb]\centering
  \input{SystemArch.tex}
  \caption{System architecture for \MYCAD~deployment to provide ubiquitous accessibility. The three
    different deployment forms (Windows DLL, Linux SO, and cloud API) are shown in red. The first
    two forms, i.e., DLL and SO, are deployed directly from compiled algorithms; both call the
    OpenCL library for speeding up computations on GPUs. The cloud-based API calls either DLL or SO
    (depending on the OS of the cloud machine) through a wrapper. The wrapper is C\# or C++ for
    windows DLL or Linux SO respectively. One use of the API will be mere direct calls by developers
    through HTTP requests. The other use of the API could be a complete cloud-based CAD (WEBCAD) for
    end-users for ubiquitous accessibility (Sec.~\ref{sec:deploying-cad-as}).}\label{fig:SystemArch}
\end{figure}
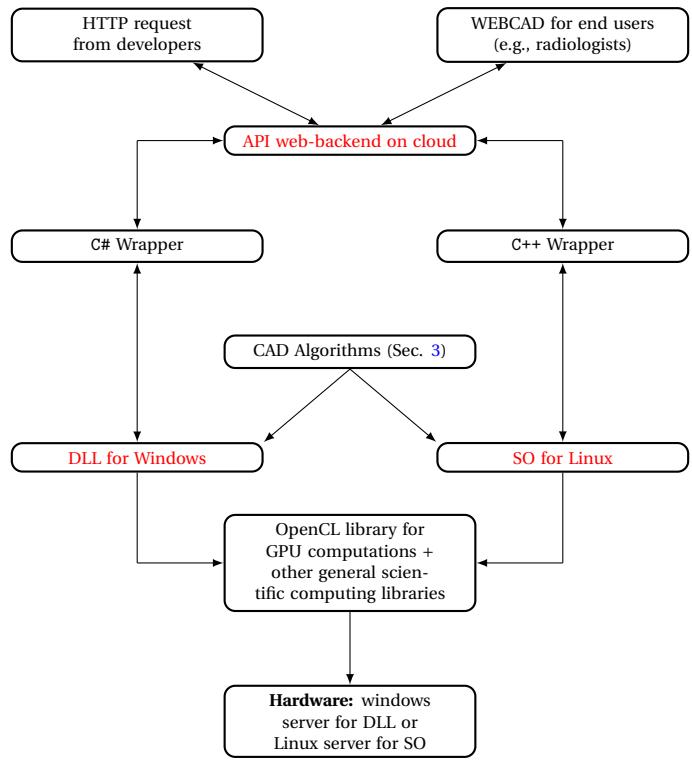
The main blocks of the system architecture are depicted in Figure~\ref{fig:SystemArch} and explained
below:
\begin{description}
  \item[DLL or SO:] all \MYCAD~algorithms are compiled to a callable library; hence the name LIBCAD
  (for LIBrary based CAD). The library is Dynamic Link Library (DLL) or Shared Object (SO) depending
  on whether the hosting machine is Windows or Linux, respectively. The importance of this step is
  that it makes the CAD functionalities independent of both the implementation and calling
  languages.

  \item [OpenCL:] This is an implementation layer of some basic and low-level computational routines
  to be able to run the DLL (or SO) on GPU. The importance of this layer is that it allows for fast
  processing necessary for some CAD functionalities, e.g., LHS of enhancement step; and allows, as
  well, for concurrent multiple users for higher level cloud-based access.

  \item[API:] This is a cloud-based Application Programming Interface (API) that is able to
  establish communication between the library (DLL or SO) and the internet external world. This API
  is implemented solely using PHP and we made it available as an opensource repository
  \citep{LIBCADUtil2012}. One way of using this API is a direct call by developers through HTTP
  requests, using any programming language, to integrate the CAD functionalities to their
  software. This software may range from simple image viewer to a sophisticated batch processing for
  research purposes. The other use of this API can be a full web-based CAD system (WEBCAD) for end
  users, e.g., radiologists.

  \item[Wrappers:] This is either C\# or C++ interface layer to the DLL or SO library
  respectively. This wrapper is necessary to communicate with the PHP layer of the cloud-based API.
\end{description}
Moving CAD to cloud using the proposed architecture (Figure~\ref{fig:SystemArch}) raises other software
engineering aspects and concerns. E.g., users (either programmers calling the API or end users using
the WEBCAD) will need to upload their mammograms to the cloud. This is in contrast to the scenario
where CADs are deployed to local mammography machines. This change of scenario comes with
challenges. For instance, the DLL (or SO) libraries require a physical path to load images and
temporarily storing them until the end of image analysis. To address that, we have utilized a
session-based storage approach, where users upload their input images only once; these images are
stored on the server for DLL (or SO) functions to access them, and are discarded after the end of
each session to conserve memory and storage. However, uploading images adds an overhead, as it
depends on the user’s internet speed which may vary in different parts of the world.

\subsection{System Assessment}\label{sec:perf-optim-hardw}
\begin{table}[tb]
  \centering
  \begin{tabular}{l l l l l}
    \toprule
    & \multicolumn{4}{c}{\bfseries Execution time:}\\
    \textbf{Hardware Configuration} & \textbf{C\#} & \textbf{DLL}& \textbf{PHP}& \textbf{Total}\\
    \midrule
    \multicolumn{5}{c}{\textbf{Image Enhancement}}\\
    2 Cores / 3.5 RAM / 2.09 GHz &9.1	&3.4	&1.1	&13.6\\
    4 Cores / 7 RAM / 2.09 GHz   &9	&3	&1.1	&13.1\\
    8 Cores / 14 RAM / 2.10 GHz  &8.8	&2.9	&0.5	&12.2\\
    \midrule
    \multicolumn{5}{c}{\textbf{Mass Detection}}\\
    2 Cores / 3.5 RAM / 2.09 GHz &9.2	&57.6	&1.4	&68.2\\
    4 Cores / 7 RAM / 2.09 GHz   &8.9	&31.7	&1.3	&41.9\\
    8 Cores / 14 RAM / 2.10 GHz  &8.9	&18.7	&1.5	&29.1\\
    \midrule
    \multicolumn{5}{c}{\textbf{Microcalcification Detection}}\\
    2 Cores / 3.5 RAM / 2.09 GHz &2.2	&16.4	&1.6	&20.2\\
    4 Cores / 7 RAM / 2.09 GHz   &2.1	&13.8	&0.9	&16.8\\
    8 Cores / 14 RAM / 2.10 GHz  &2.2	&13.4	&1.4	&17\\
    \bottomrule
  \end{tabular}
  \caption{Execution time of each block of system architecture of Figure~\ref{fig:SystemArch}
    (wrapper, DLL/SO, PHP, and total) under different hardware configurations (2, 4, 8, cores) for the
    three CAD functionalities (image enhancement, mass detection, and microcalcification
    detection).}\label{tab:execution-time-each}
\end{table}
We optimized the deployed libraries (DDL and SO), as well as, the PHP interface of the API to
utilize the GPU and the multi-core architecture of the hardware. Table~\ref{tab:execution-time-each}
is a benchmark that illustrates the execution time of each block of the architecture in
Figure~\ref{fig:SystemArch} and how the CAD utilizes the hardware architecture. It is clear that the
majority chunk of time is consumed in the heavy computations by the main major CAD functionalities
(image enhancement, mass detection, and microcalcification detection) not the wrapper nor the PHP
layer. This supports the claim of the efficiency of this system architecture.

As operations are conducted online other software engineering concerns are raised, e.g., the need
for advanced security measures become vital to ensure protection of data and transactions. The
measures include establishing secure connections and SSL certificates. These issues are out of scope
of our present article and will be discussed in another software engineering related publication.


%% file: SystemArch.tex
\if 0\MYJOURNAL%
\begin{tikzpicture}[scale=0.7]
\else \if 1\MYJOURNAL%
\begin{tikzpicture}[scale=0.5]
\fi\fi
  \node [punktchain] (a1) at ( -4,4)  {HTTP request from developers};
  \node [punktchain] (a2) at ( 4,4)  {WEBCAD for end users (e.g., radiologists)};
  \node [punktchain] (b) at ( 0,2)  {\color{red} API web-backend on cloud};
  \node [punktchain] (j) at ( 0,-2)  {CAD Algorithms (Sec. \ref{sec:algorithms})};
  \node [punktchain] (c) at ( 4,0)  {\texttt{C++} Wrapper};
  \node [punktchain] (e) at ( 4,-4)  {\color{red} SO for Linux};
  \node [punktchain] (d) at (-4,0)  {\texttt{C\#} Wrapper};
  \node [punktchain] (f) at (-4,-4)  {\color{red} DLL for Windows};
  \node [punktchain] (g) at ( 0,-6)  {OpenCL library for GPU computations $+$ other general scientific computing libraries};
  \node [punktchain] (i) at ( 0,-9)  {\textbf{Hardware:} windows server for DLL or Linux server for SO};
  \begin{scope}[>=latex]
    \draw [->] (4,2) to (c);
    \draw [->] (4,2) to (b);
    \draw [->] (-4,2) to (b);
    \draw [->] (-4,2) to (d);
    \draw [-] (4,-6) to (e);
    \draw [->] (4,-6) to (g);
    \draw [-] (-4,-6) to (f);
    \draw [->] (-4,-6) to (g);
    \draw [<->] (a1) to (b);
    \draw [<->] (a2) to (b);
    \draw [<->] (c) to (e);
    \draw [<->] (d) to (f);
    \draw [<-] (e.north west) to (j.south);
    \draw [<-] (f.north east) to (j.south);
    \draw [->] (g) to (i);
  \end{scope}
\end{tikzpicture}


%% file: SecDiscussion.tex
\section{Conclusion, Discussion, and Future Work}\label{sec:discussion}
This paper is the main publication of a project coined to design a complete Computer Aided Detection
(CAD) for detecting breast cancer from mammograms. The project involved several researchers from
several disciplines and a large database was acquired from two different institutions for design and
assessment. This CAD is complete since it is comprised of both method and system. The method is a
set of algorithms designed for image enhancement, mass detection, and microcalcification
detection. The system is an implementation and deployment of a software architecture that
facilitates both local and cloud access to the designed algorithms so that the CAD is not limited to
a particular or local mammography machine. The assessment results of both the method and the system
are presented. The results of method assessment illustrate that the CAD sensitivity is much higher
than the majority of commercial and FDA approved CADs. The results of system assessment illustrates
that the layer responsible for deploying the CAD to the cloud has a minimal effect on the CAD
performance, and that the majority of the delay comes from the algorithm computations. Phase II of
the project is just launched to consider Deep Neural Networks as the method of learning for both
mass and microcalcification detection. Phase III is still under planing which will consider
detection of other types of abnormalities. These two phases are explained next.

\subsection{Current Work: DNN and publicly available dataset}\label{sec:dnn}
As was introduced in Section~\ref{sec:review}, a very early attempt to use CNN, which belongs to the
featureless pixel-based approach, in mass and microcalcification detection was
\cite{Lo1995Artificialconvolution, Sahiner1996ClassMass,
  Chan1995MammographicMicrocalcification}. Since then, more than 20 years, the era of DNN evolved
dramatically in both directions: DNN architecture and GPU computational power. Recent attempts to
use DNN for the problem of breast cancer detection are \cite{Carneiro2017DeepLearningModelsFor,
  Akselrod-Ballin2017DeepLearningForAutomatic, Akselrod-Ballin2017CnnBasedMethodFor,
  Qiu2017NewApproachDevelopComputer, Hamidinekoo2018DeepLearningMammographyBreast,
  Dhungel2017DeepLearningApproachFor, Carneiro2017AutomatedAnalysisUnregisteredMulti,
  Dhungel2017FullyAutomatedClassificationMammograms,
  Bevilacqua2018PerformanceComparisonBetweenShallow}. We just launched a research project to
redesign the three algorithms of image enhancement, mass detection, and microcalcification detection
using DNN. We hope to come up with an architecture that wins over the designed algorithms explained
in the present article. However, it is of high interest to know whether a smart DNN architecture
would be able to beat the LHS algorithm designed for image enhancement? The motivation, explained in
Section~\ref{sec:image-enhancement}, behind the design of LHS was founded on the experience gained
from understanding the nature of mammograms. Then the previous question may be reduced to what is
the combination of the dataset size and DNN architecture that is able to win over our understanding
of the nature of mammograms. Another current interest to amend this project is to document and publish
the acquired datasets (Table~\ref{tab:datab-acqu-design}) to be publicly available for the research
community for future investigations.

\subsection{Future Work: other types of abnormalities}\label{sec:future-work}
There is a scientific debate whether the ethnicity of the training and testing datasets affect the
CAD accuracy. Provided that the CAD of the present article is trained and tested on Egyptian dataset
acquired from two different institutions, it is of great importance to us to investigate whether the
accuracy would differ when testing the CAD on other datasets with different ethnicity. A parallel
interesting study is to stratify the CAD accuracy for lesion size, which is akin to
\cite{Malich2003Influence, Brem2007ImpactOfBreastDensity}.

Another track of future research is to extend CAD capabilities to detect other types of
abnormalities or cancer, e.g., architectural distortion as pursued in \cite{Tourassi2006StudyOnComp,
  Sampat2005EvidenceArchDist, Matsubara2003AutoDetectMethod, Ichikawa2004AutoDetect,
  Guo2005InvestSVM, Ayres2005CharacArchDist} or bilateral asymmetry as pursued in
\cite{Celaya-Padilla2018ContralateralAsymmetryForBreast, Yin1991ComputerizedDetection,
  Yin1993ComparisonBilateral, Yin1994CompDetectionMasses, Yin1994ComputerizedAutomated,
  Scutt2006BreastAsym, Rangayyan2007AnaBilateralAsym, Lau1991AutoDet, Ferrari2001AnaAsym}. However,
we think that detecting this kind of abnormalities is a real challenge for DNN for that the incident
rate of this type of abnormalities in the diseased population is quite small if compared to mass or
microcalcification. Hence, there is no enough dataset for DNN to learn from, and it seems that
feature handcrafting is inevitable.


%% file: SecAcknowledgment.tex
\section{Acknowledgment}\label{sec:acknowledgment}
Many parties and individuals have contributed to this large scale project over the span of more than
5 years. The authors gratefully acknowledge the support of:
\begin{itemize}[partopsep=0in,parsep=0in,topsep=0in,itemsep=0in,leftmargin=\parindent]
  \item \textit{Information Technology Industry Development Agency} (ITIDA: \url{http://www.itida.gov.eg})
  for funding an early stage (the first year) of this project under the grant number ARP2009.R6.3.
  \item \textit{MESC for Research and Development} (\url{http://www.mesclabs.com}) for amending
  the fund of this project
  \item \textit{The National Campaign for Breast Cancer Screening} (institution A) for providing their
  large datasets of mammograms.
  \item \textit{Alfa Scan} (institution B), \url{http://www.alfascan.com.eg}, for providing their
  large datasets of mammograms.
  \item NVIDIA Corporation with the donation of the two Titan X Pascal GPU to be used for phase II
  of the project to redesign the CAD using DNN.
\end{itemize}


%% file: Algorithms.tex
\lstset{escapechar=@,style=customc}
\begin{lstlisting}[label=lst:Enhancement, caption={\texttt{EnhanceImage}: image enhancement on a single image m}]
EnhanceImage(m){
   EnhancedImage=0;
   W=81; //width of LH sliding window
   lambda=10; //Exp dist. parameter
   m=ScaleImage(m, 230); //standardize image height
   m=SegmentImage(m); //segment breast from bkgrnd
   NumPixels=BreastSize();
   for(i=0; i < NumPixels; i++)
       EnhancedImage(i) = LocalHistogramSpecification(m, i, W, Exp, lambda); // LHS (below)
   return EnhancedImage;
}
LocalHistogramSpecification(m, i, W, Exp, lambda){
    PDFwin = HistogramBins(m(i,W), 255); //initialize 256-bin hist. for window W surrounding POI i
    PDFexp = HistogramBins(Exp, lambda, 255); //create 256-bin hist. for Exp(10) dist.
    for(t=0; t<256; t++){// Calculate CDF from PDF
        CDFwin(t)=sum(PDFwin, 0, t); //Eq. (2)
        CDFexp(t)=sum(PDFexp, 0, t); //Eq. (3)
    }
    tnew=CDFexp^{-1}(CDFwin(m(i)));  //Eq. (1)
    return tnew;
}
\end{lstlisting}

\begin{lstlisting}[label=lst:Training, caption={\texttt{TrainingMassAlgorithm}: training using $M$ images}]
TrainingMassAlgorithm(TrainingDataset){
   M=length(TrainingDataset); //number of training images
   MassCentroidList   = {}; //mass centroids
   NormalCentroidList = {}; //normal centroids
   w_1=21; //size of feature window for POI i
   for(m=0; m < M; m++){
      m=EnhanceImage(m);
      MassFeatures = ConstructFeatures(m, w_1, MASS); //a NumPixelMass x 441 list to collect all pixels (NumPixelsMass) marked by radiologist as cancerours Mass and construct for each pixel the 441 features
      NormalFeatures = ConstructFeatures(m, NORMAL); //a list of NumPixelNormal x 441 to collect all pixels (NumPixelsNormal) NOT marked by radiologist as concourse and construct for each pixel the 441 features
      Append(Kmeans(MassFeatures, 100), MassCentroidList); //Kmeans on mass region to extract on ly 100 representatives and append them to MassCentroidList
      Append(Kmeans(NormalRegion, 100), NormalCentroidList); //Kmeans on normal region to extract only 100 representatives and append them to NormalCentroidList
   }
   PCVectors = PCA([NormalCentroidList, MassCentroidList], 10); //to find the first 10 principal  components of all centroids
   [NormalCentroidList, MassCentroidList] = [MassCentroidList, NormalCentroidList] * PCVectors; //projects each of the centroid list on the best 10 PCAs to produce the two final lists of
   centroids for cancer and normal, each is  100M x 10
   return MassCentroidList, NormalCentroidList, PCVectors;
}
\end{lstlisting}

\begin{lstlisting}[label=lst:Testing, caption= {\texttt{TestingMassAlgorithm}: testing on an image \texttt{m}}]
MassDetection(m){
   m = EnhanceImage(m);
   ScoreImage=zeros(size(m)); //Initialize a final score image of the same original mammogram size
   MassLocation = {}; //Initialize a final mass location list
   NumPixels = BreastSize(m);
   Features = ConstructFeatures(m, w_1, ALL); //A list of NumPixels x 441, where each row is the 441 gray level of the surrounding w_1xw_1 window
   Features = Features * PCVectors; //project the NumPixels x 441 Features on the PC vectors obtained from the training phase to produce a feature list of NumPixels x 10
   for(i=0; i < NumPixels; i++){
      ScoreImage(i) = KNN(Features(i), NormalCentroidList, MassCentroidList); //Classify the 10-dimensional feature vector of the ith POI using a KNN classifier w.r.t. the two centroids lists obtained from training phase
   }
   ScoreImage = SmoothImage(ScoreImage, 10); //Smoothing by a running 10 x 10 smoothing window
   MassLocation = FindMaxima(ScoreImage); //The locations of the potential cancers; the probability of being cancer is ScoreImage(MassLocation)
   return ScoreImage, MassLocation;
}
\end{lstlisting}

\begin{lstlisting}[label=lst:Microcalcification, caption= {\texttt{MicrocalcificationDetection} algorithm on an image \texttt{m}}]
MicrocalcificationDetection(m){
   m  = EnhanceImage(m);
   Th = 2.4;
   w_2  = 2;
   FociLocation = zeros(size(m)); //Initialize a final binary image of the same size of the original mammogram to indicate the location of microcalcification foci
   CreateFilter(InnerOuterFilter, w_2, 3 w_2); //create the filter of Figure 2
   FociLocation = FilterImage(m, InnerOuterFilter); //to select those pixels close to each others
   FociLocation = BW(FociLocation, Th); //With threshold Th, Convert FociLocation to B/W image
   FociLocation = CenterOfConnectedPixels(FociLocation); //the center of each group of connected pixels is the final focus
   return FociLocation;
}
\end{lstlisting}


%% file: Figures.tex
\begin{figure}[h]
  \centering
  \includegraphics[width=0.35\textwidth]{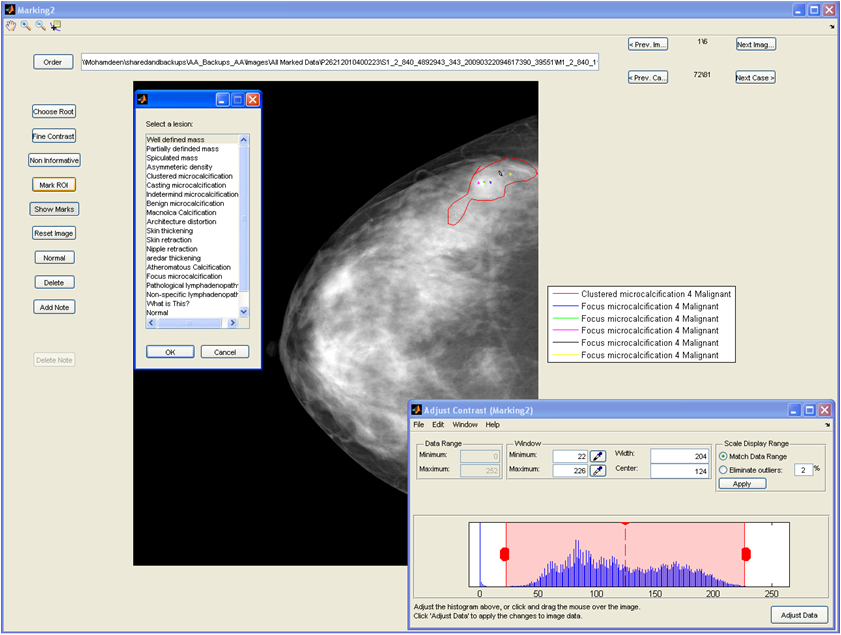}
  \caption{A snapshot for a software designed and implemented in the project to be used by the
    consultant radiologist to mark every lesion in a mammogram. The radiologist marks accurately the
    outline of a lesion with entering all related information, e.g., BIRAD and histological
    results.}\label{fig:snapsh-our-design}
\end{figure}

\if 0\MYJOURNAL%
\newcommand{\MyScale}{0.7}
\newcommand{\MYWIDTH}{0.24\textwidth}
\else \if 1\MYJOURNAL%
\newcommand{\MyScale}{0.8}
\newcommand{\MYWIDTH}{0.24\textwidth}
\fi
\fi

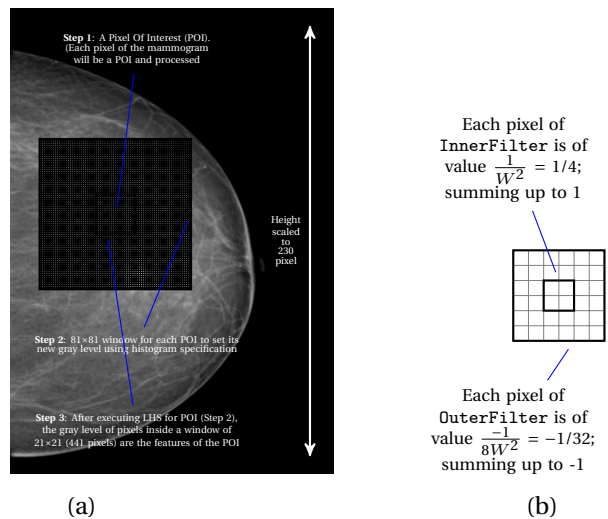
\begin{figure}[!h]
  \centering
  \begin{tikzpicture}[x=\MyScale, y=\MyScale]
    \newcommand{\MyLHis}{40}
    \newcommand{\MyLMass}{10}
    \newcommand{\MyHH}{232}
    \newcommand{\MyWW}{160}
    \newcommand{\MyCX}{-30}
    \newcommand{\MyCY}{15}

    \node (image) at (0,0) {\includegraphics[width = \MYWIDTH]{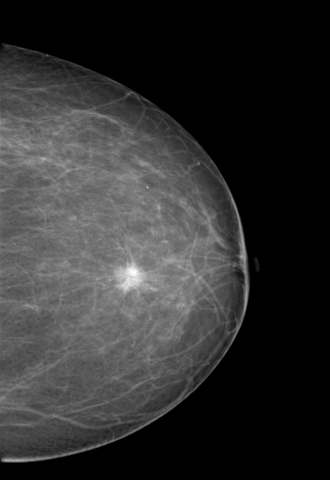}};
    \fill[blue] (\MyCX-1, \MyCY-1) rectangle(\MyCX,\MyCY);
    \draw[step=1,black,very thin] (\MyCX-\MyLHis-1, \MyCY-\MyLHis-1) grid   (\MyCX+\MyLHis, \MyCY+\MyLHis);
    \draw[thick] (\MyCX-\MyLHis-1, \MyCY-\MyLHis-1) rectangle(\MyCX+\MyLHis, \MyCY+\MyLHis);
    \draw[thick] (\MyCX-\MyLMass-1, \MyCY-\MyLMass-1) rectangle(\MyCX+\MyLMass, \MyCY+\MyLMass);

    \path (\MyCX, \MyCY-1) node(x1) {} (-0.5*\MyWW, 0.5*\MyHH) node[PIXEL, anchor=north west](y1)
    {\color{white}\textbf{Step 1}: A Pixel Of Interest (POI). (Each pixel of the mammogram will be a
      POI and processed}; \draw[-,blue] (x1) -- (y1);

    \path (\MyCX+\MyLHis, \MyCY) node(x2) {} (-0.5*\MyWW, -0.2*\MyHH) node[PIXEL, anchor=north west](y2)
    {\color{white}\textbf{Step 2}: 81$\times$81 window for each POI to set its new gray level using histogram
      specification}; \draw[-,blue] (x2) -- (y2);

    \path (\MyCX-\MyLMass+5, \MyCY-\MyLMass) node(x2) {} (-0.5*\MyWW, -0.5*\MyHH) node[PIXEL, anchor=south west](y2)
    {\color{white}\textbf{Step 3}: After executing LHS for POI (Step 2), the gray level of pixels
      inside a window of 21$\times$21 (441 pixels) are the features of the POI}; \draw[-,blue] (x2)
    -- (y2);

    \draw[step=1,color=white, <->, thick] (0.46*\MyWW, -0.5*\MyHH) -- (0.46*\MyWW, 0.5*\MyHH);
    \node[PIXEL, color=white](x2) at (0.38*\MyWW, 0){Height\\scaled\\to\\230\\pixel};
  \end{tikzpicture}%
  \begin{tikzpicture}[x=2mm, y=2mm]
    \newcommand{\W}{2}
    \draw[step=1,gray,very thin] (0,0) grid (3*\W, 3*\W);
    \draw[step=1,black,thick] (\W,\W) rectangle (2*\W, 2*\W);
    \draw[step=1,black,thick] (0,0) rectangle (3*\W, 3*\W);
    \path (1.5*\W,2*\W) node(x1) {} (0,6*\W) node[PIXEL, font=\scriptsize](y1) {Each pixel of \texttt{InnerFilter} is
      of value $\frac{1}{W^2}=1/4$; summing up to 1}; \draw[-,blue] (x1) -- (y1);
    \path (2*\W,0) node(x1) {} (0,-3*\W) node[PIXEL, font=\scriptsize](y1) {Each pixel of \texttt{OuterFilter} is
      of value $\frac{-1}{8W^2}=-1/{32}$; summing up to -1}; \draw[-,blue] (x1) -- (y1);
  \end{tikzpicture}

  (a) \hfil (b)
  \caption{The main steps of \texttt{EnhanceImage} and \texttt{MassDectection} Algorithms. (a) An
    image is reduced in size to 230-pixel height. Then the breast is segmented from the
    background. For each pixel (POI) in the breast region, a surrounding window of size 81$\times$81
    is used for local histogram specification to find the new gray level of this POI. Mass detection
    is pursued after histogram specification using another sliding window of size 21$\times$21
    centered around each POI, where its features will be a vector of the whole 21$\times$21 gray
    levels. (b) Filter of microcalcification detection is composed of two nested filters: inner of
    size $\mathtt{w2} \times \mathtt{w2}$ and outer of size $3\mathtt{w2} \times 3\mathtt{w2}$, where $\mathtt{w2}$ is chosen to be 2
    pixels.}\label{fig:wind-surr-pixel}
\end{figure}

\begin{figure}[h]\centering
  \includegraphics[width=0.24\textwidth]{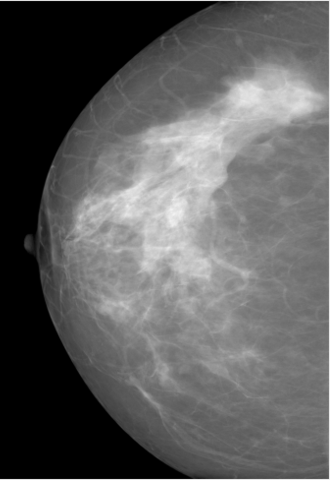}\hfil\includegraphics[width=0.24\textwidth]{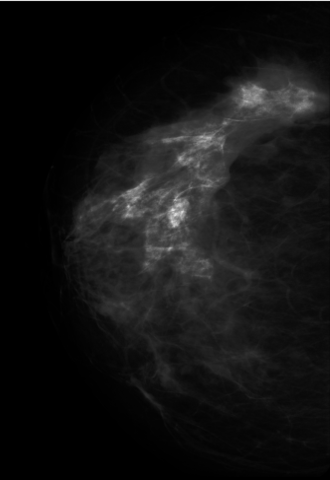}
  \caption{Image enhancement: original image (left) and the \texttt{EnhancedImage} (right) is both clearer for radiologists' inspection and standardized for algorithm detection}\label{fig:image-enhancement}
\end{figure}

\begin{figure*}
  \centering
  \includegraphics[width=0.195\textwidth]{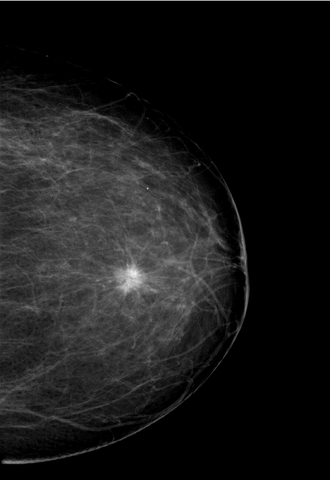}\hfil\includegraphics[width=.195\textwidth]{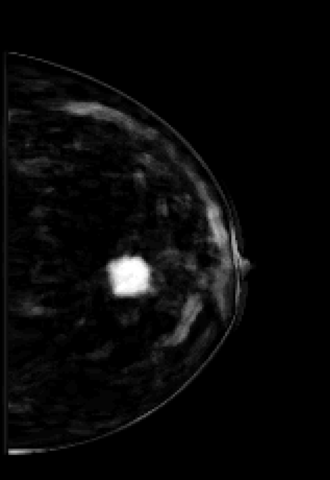}\hfil\includegraphics[width=.195\textwidth]{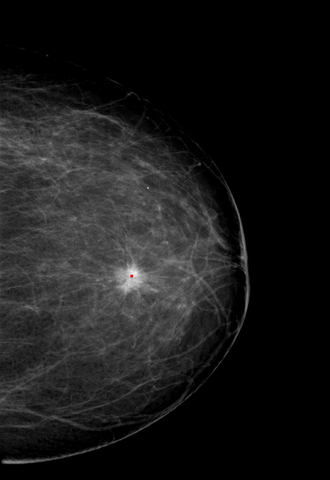}\hfil\includegraphics[width=.195\textwidth]{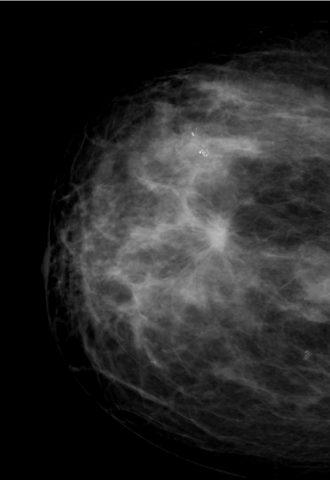}\hfil\includegraphics[width=.195\textwidth]{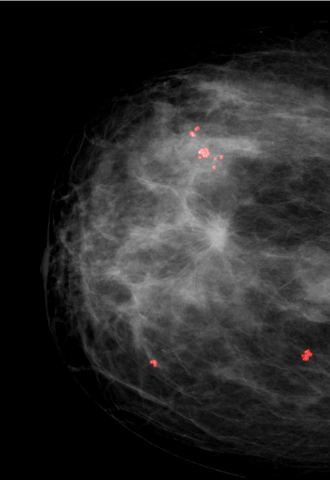}
  (a) \hspace{\TMPSPACE} (b) \hspace{\TMPSPACE} (c) \hspace{\TMPSPACE} (d) \hspace{\TMPSPACE} (e)
  \caption{Output examples for mass and microcalcification detection algorithms. (a) an image
    containing a mass after image enhancement, (b) \texttt{ScoreImage} of mass detection, (c)
    \texttt{MassLocation} plotted as a red marker on the original image, (d) an image containing
    microcalcification, and (e) \texttt{FociLocation} of the detected microcalcifications plotted as
    red points on the original image.}\label{fig:mass-micro-detect}
\end{figure*}

\begin{figure*}[h]
  \centering
  \includegraphics[width=0.195\textwidth]{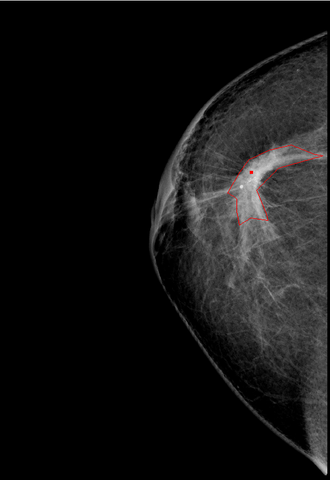}\hfil\includegraphics[width=0.195\textwidth]{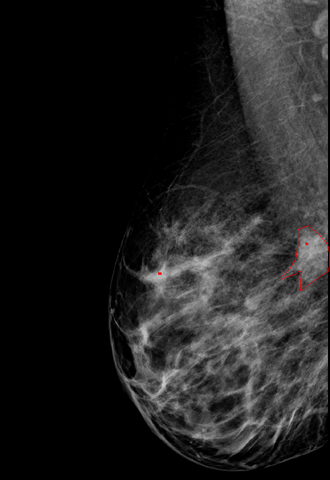}\hfil\includegraphics[width=0.195\textwidth]{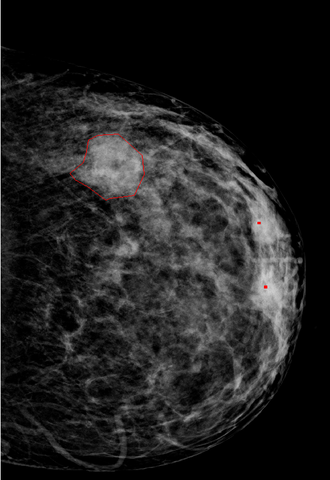}\hfil\includegraphics[width=0.195\textwidth]{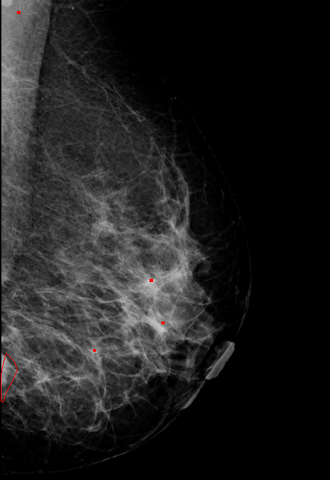}\hfil\includegraphics[width=0.195\textwidth]{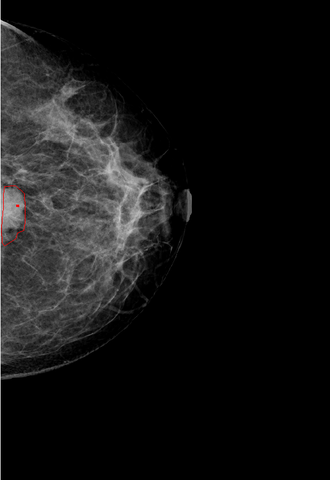}
  (a) \hspace{\TMPSPACE} (b) \hspace{\TMPSPACE} (c) \hspace{\TMPSPACE} (d) \hspace{\TMPSPACE} (e)
  \caption{Mass detection examples in different mammography views; radiologist's marking is labeled
    and CAD centroids of highest probability are indicated in red on each image. (a) mass detected
    from first centroid; (b) mass detected from second centroid with one FP; (c) mass not detected
    up to 2 FPs. (d) and (e) are two different views (MLO and CC respectively) of the same malignant
    case: (d) mass not detected up to 2 FPs; (e) mass detected from first
    centroid.}\label{fig:Mass-Centroids}
\end{figure*}


%% file: Yousef2018CAD.bbl
\begin{thebibliography}{100}
\expandafter\ifx\csname url\endcsname\relax
  \def\url#1{\texttt{#1}}\fi
\expandafter\ifx\csname urlprefix\endcsname\relax\def\urlprefix{URL }\fi
\expandafter\ifx\csname href\endcsname\relax
  \def\href#1#2{#2} \def\path#1{#1}\fi

\bibitem{Yousef2017MethodSystemForComputer}
W.~A. Yousef,
  \href{http://appft.uspto.gov/netacgi/nph-Parser?Sect1=PTO1&Sect2=HITOFF&d=PG01&p=1&u=%2Fnetahtml%2FPTO%2Fsrchnum.html&r=1&f=G&l=50&s1=%2220190019291%22.PGNR.&OS=DN/20190019291&RS=DN/20190019291}{Method
  and system for image analysis to detect cancer}, patent pending, US
  62/531,219 (07 2017).

\bibitem{Althuis2005GlobalTrendsBreastCancer}
M.~D. Althuis, J.~M. Dozier, W.~F. Anderson, S.~S. Devesa, L.~A. Brinton,
  \href{https://doi.org/10.1093/ije/dyh414}{{Global Trends in Breast Cancer
  Incidence and Mortality 1973-1997}}, Int. J. Epidemiol. 34~(2) (2005)
  405--412 (2005).

\bibitem{Freedman2006CancerIncidenceMECC}
L.~S. Freedman, B.~K. Edwards, L.~A.~G. Ries, J.~L. Young, {Cancer Incidence in
  Four Member Countries (Cyprus, Egypt, Israel, and Jordan) of the Middle East
  Cancer Consortium (MECC) Compared with US SEER.} (2006).

\bibitem{Tabar1999NaturalHistCarcinoma}
L.~Tabar, S.~W. Duffy, B.~Vitak, H.~H. Chen, T.~C. Prevost, {The Natural
  History of Breast Carcinoma: What Have We Learned From screening?}, Cancer
  86~(3) (1999) 449--462 (1999).

\bibitem{Anderson2000NaturalHistCarcinoma}
T.~J. Anderson, F.~E. Alexander, P.~M. Forrest, {The Natural History of Breast
  Carcinoma: What Have We Learned From screening?}, Cancer 88~(7) (2000)
  1758--1759 (2000).

\bibitem{Ferlay2004CancerIncidence}
J.~Ferlay, F.~Bray, P.~Pisani, D.~M. Parkin, {GLOBOCAN 2002: Cancer Incidence,
  Mortality and Prevalence Worldwide.} (2004).

\bibitem{Parkin2005GlobalCancer}
D.~M. Parkin, F.~Bray, J.~Ferlay, P.~Pisani, {Global Cancer Statistics, 2002},
  Ca-a Cancer Journal for Clinicians 55~(2) (2005) 74--108 (2005).

\bibitem{Kamal2007MissedBreast}
R.~M. Kamal, N.~M. {Abdel Razek}, M.~A. Hassan, M.~A. Shaalan, {Missed Breast
  Carcinoma; Why and How To avoid?}, J Egypt Natl Canc Inst 19~(3) (2007)
  178--194 (2007).

\bibitem{Radhika2005ImagingTechniques}
S.~Radhika, {Imaging Techniques Alternative To Mammography for Early Detection
  of Breast cancer}, TRCT (2005) 1--120 (2005).

\bibitem{Wagner2002AssessmentOf}
R.~F. Wagner, S.~V. Beiden, G.~Campbell, C.~E. Metz, W.~M. Sacks, {Assessment
  of Medical Imaging and Computer-Assist Systems: Lesson From Recent
  Experience}, Acad Radiol 9 (2002) 1264--1277 (2002).

\bibitem{Muttarak2006BreastCarcinomas}
M.~Muttarak, S.~Pojchamarnwiputh, B.~Chaiwun, {Breast Carcinomas: Why Are They
  missed?}, Singapore Med J 47~(10) (2006) 851--857 (2006).

\bibitem{RaoHowWidelyCAD2010}
V.~M. Rao, D.~C. Levin, L.~Parker, B.~Cavanaugh, A.~J. Frangos, J.~H. Sunshine,
  \href{https://doi.org/10.1016/j.jacr.2010.05.019}{{How Widely Is
  Computer-Aided Detection Used in Screening and Diagnostic mammography?}},
  Journal of the American College of Radiology : JACR 7~(10) (2010) 802--5 (oct
  2010).

\bibitem{Yousef2010OnDetecting}
W.~a. Yousef, W.~a. Mustafa, A.~a. Ali, N.~a. Abdelrazek, A.~M. Farrag,
  \href{https://doi.org/10.1109/AIPR.2010.5759684}{{On Detecting Abnormalities
  in Digital mammography}}, 2010 IEEE 39th Applied Imagery Pattern Recognition
  Workshop (AIPR) (2010) 1--7 (oct 2010).

\bibitem{AbdelRazek2012MicroclacificationLIBCAD}
N.~M. {Abdel Razek}, W.~A. Yousef, W.~A. Mustafa,
  \href{https://doi.org/10.1594/ecr2012/C-1063}{Microcalcification detection
  with and without prototype cad system (libcad): a comparative study}, in:
  European Society of Radiology, ECR 2012 / C-1063, 2012, pp. 1--15 (2012).

\bibitem{AbdelRazek2013MicroclacificationLIBCAD}
N.~M. {Abdel Razek}, W.~A. Yousef, W.~A. Mustafa, {Microcalcification Detection
  With and Without Cad System (LIBCAD): a Comparative study}, The Egyptian
  Journal of Radiology and Nuclear Medicine 44~(2) (2013) 397--404 (2013).

\bibitem{Jalalian2017FoundationMethodologiesComputerAided}
A.~Jalalian, S.~Mashohor, R.~Mahmud, B.~Karasfi, M.~I.~B. Saripan, A.~R.~B.
  Ramli, Foundation and methodologies in computer-aided diagnosis systems for
  breast cancer detection, EXCLI journal 16 (2017) 113 (2017).

\bibitem{Zhao2018MinimizationAnnotationWork}
Y.~Zhao, J.~Zhang, H.~Xie, S.~Zhang, L.~Gu, Minimization of annotation work:
  Diagnosis of mammographic masses via active learning, Physics in Medicine \&
  Biology 63~(11) (2018) 115003 (2018).

\bibitem{Zhao2018MammographicImageClassificationSystem}
Y.~Zhao, D.~Chen, H.~Xie, S.~Zhang, L.~Gu, Mammographic image classification
  system via active learning, Journal of Medical and Biological Engineering
  (2018) 1--14 (2018).

\bibitem{Cheng2003CADsurveyMC}
H.~D. Cheng, X.~P. Cai, X.~W. Chen, L.~M. Hu, X.~L. Lou,
  \href{https://doi.org/Doi 10.1016/S0031-3203(03)00192-4}{{Computer-Aided
  Detection and Classification of Microcalcifications in Mammograms: a
  survey}}, Pattern Recognition 36~(12) (2003) 2967--2991 (2003).

\bibitem{Cheng2006ApproachesAutoDetClass}
H.~D. Cheng, X.~J. Shi, R.~Min, L.~M. Hu, X.~P. Cai, H.~N. Du, {Approaches for
  Automated Detection and Classification of Masses in mammograms}, Pattern
  Recognition 39~(4) (2006) 646 (2006).

\bibitem{Bozek2009SurveyImageProc}
J.~Bozek, M.~Mustra, K.~Delac, M.~Grgic,
  \href{https://doi.org/10.1007/978-3-642-02900-4_24}{{A Survey of Image
  Processing Algorithms in Digital Mammography}}  631--657
\bibitem{Tang2009CADreview}
J.~S. Tang, R.~M. Rangayyan, J.~Xu, I.~{El Naqa}, Y.~Y. Yang,
  \href{https://doi.org/10.1109/titb.2008.2009441}{{Computer-Aided Detection
  and Diagnosis of Breast Cancer With Mammography: Recent Advances}}, IEEE
  Transactions on Information Technology in Biomedicine 13~(2) (2009) 236--251
  (2009).

\bibitem{Rangayyan2007ReviewofCAD}
R.~M. Rangayyan, F.~J. Ayres, J.~E.~L. Desautels, \href{https://doi.org/DOI
  10.1016/j.jfranklin.2006.09.003}{{A Review of Computer-Aided Diagnosis of
  Breast Cancer: Toward the Detection of Subtle signs}}, Journal of the
  Franklin Institute-Engineering and Applied Mathematics 344~(3-4) (2007)
  312--348 (2007).

\bibitem{Oliver2010ReviewMassDet}
A.~Oliver, J.~Freixenet, J.~Marti, E.~Perez, J.~Pont, E.~R.~E. Denton,
  R.~Zwiggelaar, \href{https://doi.org/10.1016/j.media.2009.12.005}{{A Review
  of Automatic Mass Detection and Segmentation in Mammographic images}},
  Medical Image Analysis 14~(2) (2010) 87--110 (2010).

\bibitem{Adler1995NewMethodsForImaging}
D.~D. Adler, R.~L. Wahl, {NEW Methods for Imaging the Breast - Techniques,
  Findings, and POTENTIAL}, American Journal of Roentgenology 164~(1) (1995)
  19--30 (1995).

\bibitem{Giger1996b}
M.~Giger, H.~MacMahon, {Image Processing and Computer-Aided diagnosis},
  Radiologic Clinics of North America 34~(3) (1996) 565--+ (1996).

\bibitem{Vyborny2000CAD}
C.~J. Vyborny, M.~L. Giger, R.~M. Nishikawa, {Computer-Aided Detection and
  Diagnosis of Breast cancer}, Radiologic Clinics of North America 38~(4)
  (2000) 725--+ (2000).

\bibitem{Singh2008b}
V.~Singh, C.~Saunders, L.~Wylie, A.~Bourke,
  \href{https://doi.org/10.2217/14796694.4.4.501}{{New Diagnostic Techniques
  for Breast Cancer detection}}, Future Oncology 4~(4) (2008) 501--513 (2008).

\bibitem{Giger2008b}
M.~L. Giger, H.~P. Chan, J.~Boone,
  \href{https://doi.org/10.1118/1.3013555}{{Anniversary Paper: History and
  Status of Cad and Quantitative Image Analysis: the Role of Medical Physics
  and AAPM}}, Medical Physics 35~(12) (2008) 5799--5820 (2008).

\bibitem{Elter2009CADx}
M.~Elter, A.~Horsch, {CADx of Mammographic Masses and Clustered
  Microcalcifications: a review}, Med Phys 36~(6) (2009) 2052--2068 (2009).

\bibitem{Otsu1979ThresholdSelectionMethodFrom}
N.~Otsu, A threshold selection method from gray-level histograms, IEEE
  transactions on systems, man, and cybernetics 9~(1) (1979) 62--66 (1979).

\bibitem{Ojala2001AccurateSegmentation}
T.~Ojala, J.~Nappi, O.~Nevalainen, {Accurate Segmentation of the Breast Region
  From Digitized mammograms}, Computerized Medical Imaging and Graphics 25~(1)
  (2001) 47--59 (2001).

\bibitem{Ferrari2004IdentificationBreastBoundaryMammograms}
R.~J. Ferrari, A.~F. Fr{\`e}re, P.~R.~M. Rangayyan, J.~E.~L. Desautels, R.~A.
  Borges, Identification of the breast boundary in mammograms using active
  contour models, Medical and Biological Engineering and Computing 42 (2004)
  201--208 (2004).

\bibitem{Sivaramakrishna2000ComparingPerformance}
R.~Sivaramakrishna, N.~A. Obuchowski, W.~A. Chilcote, G.~Cardenosa, K.~A.
  Powell, {Comparing the Performance of Mammographic Enhancement Algorithms: a
  Preference study}, American Journal of Roentgenology 175~(1) (2000) 45--51
  (2000).

\bibitem{Singh2005AnEvaluation}
S.~Singh, K.~Bovis, \href{https://doi.org/10.1109/titb.2004.837581}{{An
  Evaluation of Contrast Enhancement Techniques for Mammographic Breast
  masses}}, IEEE Transactions on Information Technology in Biomedicine 9~(1)
  (2005) 109--119 (2005).

\bibitem{Sakellaropoulos2003WaveletBasedSpatially}
P.~Sakellaropoulos, L.~Costaridou, G.~Panayiotakis, \href{https://doi.org/Pii
  s0031-9155(03)53455-8 10.1088/0031-9155/48/6/307}{{A Wavelet-Based Spatially
  Adaptive Method for Mammographic Contrast enhancement}}, Physics in Medicine
  and Biology 48~(6) (2003) 787--803 (2003).

\bibitem{Scharcanski2006DenoisiningEnhancing}
J.~Scharcanski, C.~R. Jung,
  \href{https://doi.org/10.1016/j.compmedimag.2006.05.002}{{Denoising and
  Enhancing Digital Mammographic Images for Visual screening}}, Computerized
  Medical Imaging and Graphics 30~(4) (2006) 243--254 (2006).

\bibitem{Tang2007ImageEnhancementAlgorithm}
J.~Tang, S.~Qingling, K.~Agyepong,
  \href{https://doi.org/10.1109/icip.2007.4379757}{{An Image Enhancement
  Algorithm Based on a Contrast Measure in the Wavelet Domain for Screening
  Mammograms}}, in: Image Processing, 2007. ICIP 2007. IEEE International
  Conference on, Vol.~5, 2007, pp. V -- 29--V -- 32 (2007).

\bibitem{Dhawan1986EnhancementMammoFeatures}
A.~P. Dhawan, G.~Buelloni, R.~Gordon,
  \href{https://doi.org/10.1109/tmi.1986.4307733}{{Enhancement of Mammographic
  Features By Optimal Adaptive Neighborhood Image Processing}}, Medical
  Imaging, IEEE Transactions on 5~(1) (1986) 8--15 (1986).

\bibitem{Morrow1992RegionBasedContrast}
W.~M. Morrow, R.~B. Paranjape, R.~M. Rangayyan, J.~E.~L. Desautels,
  \href{https://doi.org/10.1109/42.158944}{{REGION-BASED Contrast Enhancement
  of MAMMOGRAMS}}, IEEE Transactions on Medical Imaging 11~(3) (1992) 392--406
  (1992).

\bibitem{Petrick1995Automated}
N.~Petrick, H.~P. Chan, B.~Sahiner, D.~T. Wei, M.~A. Helvie, M.~M. Goodsitt,
  D.~D. Adler, {Automated Detection of Breast Masses on Digital Mammograms
  Using Adaptive Density-Weighted Contrast Enhancement Filtering}, Medical
  Imaging 1995: Image Processing 2434 (1995) 590--597 (1995).

\bibitem{Petrick1996AdaptiveDensityWeighted}
N.~Petrick, H.~P. Chan, B.~Sahiner, D.~T. Wei,
  \href{https://doi.org/10.1109/42.481441}{{An Adaptive Density-Weighted
  Contrast Enhancement Filter for Mammographic Breast Mass detection}}, IEEE
  Transactions on Medical Imaging 15~(1) (1996) 59--67 (1996).

\bibitem{Rangayyan1997ImprovementSensitivity}
R.~M. Rangayyan, L.~Shen, Y.~Shen, J.~E. Desautels, H.~Bryant, T.~J. Terry,
  N.~Horeczko, M.~S. Rose,
  \href{https://doi.org/10.1109/4233.654859}{{Improvement of Sensitivity of
  Breast Cancer Diagnosis With Adaptive Neighborhood Contrast Enhancement of
  mammograms}}, IEEE transactions on information technology in biomedicine : a
  publication of the IEEE Engineering in Medicine and Biology Society 1~(3)
  (1997) 161--170 (1997).

\bibitem{Kim1997Adaptivemammographicimage}
J.~K. Kim, J.~M. Park, K.~S. Song, H.~W. Park, {Adaptive Mammographic Image
  Enhancement Using First Derivative and Local statistics}, IEEE Transactions
  on Medical Imaging 16~(5) (1997) 495--502 (1997).

\bibitem{Veldkamp2000NormalizationMammo}
W.~J. Veldkamp, N.~Karssemeijer,
  \href{https://doi.org/10.1109/42.875197}{{Normalization of Local Contrast in
  mammograms}}, IEEE Trans Med Imaging 19~(7) (2000) 731--738 (2000).

\bibitem{Gupta1995TheUseofTexture}
R.~Gupta, P.~E. Undrill, {The Use of Texture Analysis To Delineate Suspicious
  Masses in Mammography}, Physics in Medicine and Biology 40~(5) (1995)
  835--855 (1995).

\bibitem{McLoughlin2004NoiseEqualization}
K.~J. McLoughlin, P.~J. Bones, N.~Karssemeijer, \href{https://doi.org/Doi
  10.1109/Tmi.2004.824240}{{Noise Equalization for Detection of
  Microcalcification Clusters in Direct Digital Mammogram images}}, IEEE
  Transactions on Medical Imaging 23~(3) (2004) 313--320 (2004).

\bibitem{Petrick1996Automated}
N.~Petrick, H.~P. Chan, D.~T. Wei, B.~Sahiner, M.~A. Helvie, D.~D. Adler,
  {Automated Detection of Breast Masses on Mammograms Using Adaptive Contrast
  Enhancement and Texture classification}, Medical Physics 23~(10) (1996)
  1685--1696 (1996).

\bibitem{Petrick1999CombAdap}
N.~Petrick, H.~P. Chan, B.~Sahiner, M.~A. Helvie, {Combined Adaptive
  Enhancement and Region-Growing Segmentation of Breast Masses on Digitized
  mammograms}, Med Phys 26~(8) (1999) 1642--1654 (1999).

\bibitem{Ciecholewski2017MalignantBenignMassSegmentation}
M.~Ciecholewski, Malignant and benign mass segmentation in mammograms using
  active contour methods, Symmetry 9~(11) (2017) 277 (2017).

\bibitem{Rangayyan1997MeasuresAcut}
R.~M. Rangayyan, N.~M. El-Faramawy, J.~E. Desautels, O.~A. Alim,
  \href{https://doi.org/10.1109/42.650876}{{Measures of Acutance and Shape for
  Classification of Breast tumors}}, IEEE Trans Med Imaging 16~(6) (1997)
  799--810 (1997).

\bibitem{Sahiner2001Improvement}
B.~Sahiner, H.~P. Chan, N.~Petrick, M.~A. Helvie, L.~M. Hadjiiski, {Improvement
  of Mammographic Mass Characterization Using Spiculation Meausures and
  Morphological features}, Med Phys 28~(7) (2001) 1455--1465 (2001).

\bibitem{Ojala1996ComparativeStudy}
T.~Ojala, M.~Pietikainen, D.~Harwood, {A Comparative Study of Texture Measures
  With Classification Based on Feature distributions}, Pattern Recognition
  29~(1) (1996) 51--59 (1996).

\bibitem{Harwood1995TextureClassification}
D.~Harwood, T.~Ojala, M.~Pietikainen, S.~Kelman, L.~Davis, {Texture
  Classification By Center-Symmetrical Autocorrelation, Using Kullback
  Discrimination of Distributions}, Pattern Recognition Letters 16~(1) (1995)
  1--10 (1995).

\bibitem{Chan1995CAD}
H.~P. Chan, D.~Wei, M.~A. Helvie, B.~Sahiner, D.~D. Adler, M.~M. Goodsitt,
  N.~Petrick, {Computer-Aided Classification of Mammographic Masses and Normal
  Tissue: Linear Discriminant Analysis in Texture Feature space}, Phys Med Biol
  40~(5) (1995) 857--876 (1995).

\bibitem{Sahiner1998Computerized}
B.~Sahiner, H.~P. Chan, N.~Petrick, M.~A. Helvie, M.~M. Goodsitt, {Computerized
  Characterization of Masses on Mammograms: the Rubber Band Straightening
  Transform and Texture analysis}, Med Phys 25~(4) (1998) 516--526 (1998).

\bibitem{Karssemeijer1996Detection}
N.~Karssemeijer, G.~M. {Te Brake},
  \href{https://doi.org/10.1109/42.538938}{{Detection of Stellate Distortions
  in mammograms}}, IEEE Trans Med Imaging 15~(5) (1996) 611--619 (1996).

\bibitem{Kobatake1999ConvergenceIndx}
H.~Kobatake, S.~Hashimoto, {Convergence Index Filter for Vector fields}, Ieee
  Transactions on Image Processing 8~(8) (1999) 1029--1038 (1999).

\bibitem{Kobatake1999CompDetec}
H.~Kobatake, M.~Murakami, H.~Takeo, S.~Nawano,
  \href{https://doi.org/10.1109/42.774164}{{Computerized Detection of Malignant
  Tumors on Digital mammograms}}, IEEE Trans Med Imaging 18~(5) (1999) 369--378
  (1999).

\bibitem{Varela2007Computerized}
C.~Varela, P.~G. Tahoces, A.~J. Mendez, M.~Souto, J.~J. Vidal,
  \href{https://doi.org/DOI 10.1016/j.compbiomed.2005.12.006}{{Computerized
  Detection of Breast Masses in Digitized mammograms}}, Computers in Biology
  and Medicine 37~(2) (2007) 214--226 (2007).

\bibitem{Midya2018EdgeWeightedLocalTexture}
A.~Midya, R.~Rabidas, A.~Sadhu, J.~Chakraborty, Edge weighted local texture
  features for the categorization of mammographic masses, Journal of Medical
  and Biological Engineering (2018) 1--12 (2018).

\bibitem{Shastri2018DensityWiseTwoStage}
A.~A. Shastri, D.~Tamrakar, K.~Ahuja, Density-wise two stage mammogram
  classification using texture exploiting descriptors, Expert Systems with
  Applications 99 (2018) 71--82 (2018).

\bibitem{Zwiggelaar1999ModelBased}
R.~Zwiggelaar, T.~C. Parr, J.~E. Schumm, I.~W. Hutt, C.~J. Taylor, S.~M.
  Astley, C.~R. Boggis, \href{https://doi.org/S1361-8415(99)80016-4
  [pii]}{{Model-Based Detection of Spiculated Lesions in mammograms}}, Med
  Image Anal 3~(1) (1999) 39--62 (1999).

\bibitem{Zwiggelaar1999DetectionMass}
R.~Zwiggelaar, C.~J. Taylor, C.~M.~E. Rubin, {Detection of the Central Mass of
  Spiculated Lesions - Signature Normalisation and Model Data aspects},
  Information Processing in Medical Imaging, Proceedings 1613 (1999) 406--411
  (1999).

\bibitem{Zwiggelaar2004LinearStruc}
R.~Zwiggelaar, S.~M. Astley, C.~R.~M. Boggis, C.~J. Taylor,
  \href{https://doi.org/10.1109/tmi.2004.828675}{{Linear Structures in
  Mammographic Images: Detection and classification}}, IEEE Transactions on
  Medical Imaging 23~(9) (2004) 1077--1086 (2004).

\bibitem{Hastie1999StatMeasureCAD}
T.~Hastie, D.~Ikeda, R.~Tibshirani, {Statistical Measures for the
  Computer-Aided Diagnosis of Mammographic masses}, Journal of Computational
  and Graphical Statistics 8~(3) (1999) 531--543 (1999).

\bibitem{Li1997DigitalMammography}
L.~H. Li, W.~Qian, L.~P. Clarke, {Digital Mammography: Computer-Assisted
  Diagnosis Method for Mass Detection With Multiorientation and Multiresolution
  Wavelet transforms}, Academic Radiology 4~(11) (1997) 724--731 (1997).

\bibitem{Qian1999ImageFeature}
W.~Qian, L.~H. Li, L.~P. Clarke, {Image Feature Extraction for Mass Detection
  in Digital Mammography: Influence of Wavelet analysis}, Medical Physics
  26~(3) (1999) 402--408 (1999).

\bibitem{Qian2001DigitalMammography}
W.~Qian, X.~J. Sun, D.~S. Song, R.~A. Clark, {Digital Mammography: Wavelet
  Transform and Kalman-Filtering Neural Network in Mass Segmentation and
  detection}, Academic Radiology 8~(11) (2001) 1074--1082 (2001).

\bibitem{Li2002ComputerAided}
L.~H. Li, R.~A. Clark, J.~A. Thomas, {Computer-Aided Diagnosis of Masses With
  Full-Field Digital mammography}, Academic Radiology 9~(1) (2002) 4--12
  (2002).

\bibitem{Rashed2007Multiresolution}
E.~A. Rashed, I.~A. Ismail, S.~I. Zaki, \href{https://doi.org/DOI
  10.1016/j.patrec.2006.07.010}{{Multiresolution Mammogram Analysis in
  Multilevel decomposition}}, Pattern Recognition Letters 28~(2) (2007)
  286--292 (2007).

\bibitem{Anitha2017DualStageAdaptiveThresholding}
J.~Anitha, J.~D. Peter, S.~I.~A. Pandian, A dual stage adaptive thresholding
  (dusat) for automatic mass detection in mammograms, Computer methods and
  programs in biomedicine 138 (2017) 93--104 (2017).

\bibitem{Chakraborty2018ComputerAidedDetectionDiagnosis}
J.~Chakraborty, A.~Midya, R.~Rabidas, Computer-aided detection and diagnosis of
  mammographic masses using multi-resolution analysis of oriented tissue
  patterns, Expert Systems with Applications 99 (2018) 168--179 (2018).

\bibitem{Chakraborty2018ComputerAidedDetectionMammographic}
J.~Chakraborty, A.~Midya, S.~Mukhopadhyay, R.~M. Rangayyan, A.~Sadhu,
  V.~Singla, N.~Khandelwal, Computer-aided detection of mammographic masses
  using hybrid region growing controlled by multilevel thresholding, Journal of
  Medical and Biological Engineering (2018) 1--15 (2018).

\bibitem{Lo1995Artificialconvolution}
S.~C.~B. Lo, H.~P. Chan, J.~S. Lin, H.~Li, M.~T. Freedman, S.~K. Mun,
  {Artificial Convolution Neural Network for Medical Image Pattern
  recognition}, Neural Networks 8~(7-8) (1995) 1201--1214 (1995).

\bibitem{Sahiner1996ClassMass}
B.~Sahiner, H.~P. Chan, N.~Petrick, D.~Wei, M.~A. Helvie, D.~D. Adler, M.~M.
  Goodsitt, \href{https://doi.org/10.1109/42.538937}{{Classification of Mass
  and Normal Breast Tissue: a Convolution Neural Network Classifier With
  Spatial Domain and Texture images}}, IEEE Trans Med Imaging 15~(5) (1996)
  598--610 (1996).

\bibitem{Lai1989OnTechniques}
S.~M. Lai, X.~B. Li, W.~F. Bischof, {On Techniques for Detecting Circumscribed
  Masses in Mammograms}, IEEE Transactions on Medical Imaging 8~(4) (1989)
  377--386 (1989).

\bibitem{Brake1999SingleMultiDetMammo}
G.~M. te~Brake, N.~Karssemeijer,
  \href{https://doi.org/10.1109/42.790462}{{Single and Multiscale Detection of
  Masses in Digital mammograms}}, IEEE Trans Med Imaging 18~(7) (1999) 628--639
  (1999).

\bibitem{Freixenet2008Eigendetection}
J.~Freixenet, A.~Oliver, R.~Marti, X.~Llado, J.~Pont, E.~Perez, E.~R.~E.
  Denton, R.~Zwiggelaar,
  \href{https://doi.org/10.1118/1.2897950}{{Eigendetection of Masses
  Considering False Positive Reduction and Breast Density information}},
  Medical Physics 35~(5) (2008) 1840--1853 (2008).

\bibitem{Kegelmeyer1994CompAidMammo}
W.~P. {Kegelmeyer Jr}, J.~M. Pruneda, P.~D. Bourland, A.~Hillis, M.~W. Riggs,
  M.~L. Nipper, {Computer-Aided Mammographic Screening for Spiculated lesions},
  Radiology 191~(2) (1994) 331--337 (1994).

\bibitem{Campanini2004Featureless}
R.~Campanini, D.~Dongiovanni, E.~Iampieri, N.~Lanconelli, M.~Masotti,
  G.~Palermo, A.~Riccardi, M.~Roffilli, \href{https://doi.org/Doi
  10.1088/0031-9155/49/6/007 Pii S0031-9155(04)69324-9}{{A Novel Featureless
  Approach To Mass Detection in Digital Mammograms Based on Support Vector
  machines}}, Physics in Medicine and Biology 49~(6) (2004) 961--975 (2004).

\bibitem{Choi2014CADDetectionEnsemble}
J.~Y. Choi, D.~H. Kim, K.~N. Plataniotis, Y.~M. Ro,
  \href{https://doi.org/10.1088/0031-9155/59/14/3697}{{Computer-Aided Detection
  (cad) of Breast Masses in Mammography: Combined Detection and Ensemble
  classification.}}, Physics in medicine and biology 59~(14) (2014) 3697--719
  (2014).

\bibitem{Qian1993TreeStruc}
W.~Qian, L.~P. Clarke, M.~Kallergi, H.~D. Li, R.~Velthuizen, R.~A. Clark, M.~L.
  Silbiger, {Tree-Structured Nonlinear Filter and Wavelet Transform for
  Microcalcification Segmentation in Mammography}, Biomedical Image Processing
  and Biomedical Visualization, Pts 1 and 2 1905 (1993) 509--520 (1993).

\bibitem{Yoshida1995OptimizingWavelet}
H.~Yoshida, W.~Zhang, W.~D. Cai, K.~N. Doi, R.~M. Nishikawa, M.~L. Giger,
  {Optimizing Wavelet Transform Based on Supervised Learning for Detection of
  Micro Calcifications in Digital mammograms}, International Conference on
  Image Processing - Proceedings, Vols I-Iii (1995) C152--C155 2002 (1995).

\bibitem{Strickland1996Wavelet}
R.~N. Strickland, H.~I. Hahn, {Wavelet Transforms for Detecting
  Microcalcifications in mammograms}, IEEE Transactions on Medical Imaging
  15~(2) (1996) 218--229 (1996).

\bibitem{Chen1997OnDigital}
C.~H. Chen, G.~G. Lee, {On Digital Mammogram Segmentation and
  Microcalcification Detection Using Multiresolution Wavelet Analysis},
  Graphical Models and Image Processing 59~(5) (1997) 349--364 (1997).

\bibitem{Yu2000CADsystem}
S.~Y. Yu, L.~Guan, {A Cad System for the Automatic Detection of Clustered
  Microcalcifications in Digitized Mammogram films}, IEEE Transactions on
  Medical Imaging 19~(2) (2000) 115--126 (2000).

\bibitem{Heinlein2006IntegratedWavelets}
P.~Heinlein, J.~Drexl, W.~Schneider,
  \href{https://doi.org/10.1109/tmi.2003.809632}{{Integrated Wavelets for
  Enhancement of Microcalcifications in Digital mammography}}, IEEE
  Transactions on Medical Imaging 22~(3) (2003) 402--413 (2003).

\bibitem{Karahaliou2008BreastCancerDiag}
A.~N. Karahaliou, I.~S. Boniatis, S.~G. Skiadopoulos, F.~N. Sakellaropoulos,
  N.~S. Arikidis, E.~A. Likaki, G.~S. Panayiotakis, L.~I. Costaridou,
  \href{https://doi.org/Doi 10.1109/Titb.2008.920634}{{Breast Cancer Diagnosis:
  Analyzing Texture of Tissue Surrounding Microcalcifications}}, IEEE
  Transactions on Information Technology in Biomedicine 12~(6) (2008) 731--738
  (2008).

\bibitem{Singh2006SVMbased}
S.~Singh, V.~Kumar, H.~K. Verma, D.~Singh,
  \href{https://doi.org/10.1109/IEMBS.2006.259320}{{SVM Based System for
  Classification of Microcalcifications in Digital mammograms}}, Conf Proc IEEE
  Eng Med Biol Soc 1 (2006) 4747--4750 (2006).

\bibitem{Nishikawa1995CAD}
R.~M. Nishikawa, M.~L. Giger, K.~Doi, C.~J. Vyborny, R.~A. Schmidt,
  {Computer-Aided Detection of Clustered Microcalcifications on Digital
  mammograms}, Med Biol Eng Comput 33~(2) (1995) 174--178 (1995).

\bibitem{Ge2008VascularCalcification}
J.~Ge, H.~P. Chan, B.~Sahiner, C.~Zhou, M.~A. Helvie, J.~Wei, L.~M. Hadjiiski,
  Y.~Zhang, Y.~T. Wu, J.~Shi, \href{https://doi.org/Doi
  10.1117/12.773096}{{Automated Detection of Breast Vascular Calcification on
  Full-Field Digital Mammograms - Art. No. 691517}}, Medical Imaging 2008:
  Computer-Aided Diagnosis, Pts 1 and 2 6915 (2008) 91517 (2008).

\bibitem{Karssemeijer1992StochasticModel}
N.~Karssemeijer, {Stochastic-Model for Automated Detection of Calcifications in
  Digital Mammograms}, Image and Vision Computing 10~(6) (1992) 369--375
  (1992).

\bibitem{Chan1987ImageFeatureAnaMicrocalc}
H.~P. Chan, K.~Doi, S.~Galhotra, C.~J. Vyborny, H.~Macmahon, P.~M. Jokich,
  \href{https://doi.org/Doi 10.1118/1.596065}{{Image Feature Analysis and
  Computer-Aided Diagnosis in Digital Radiography .1. Automated Detection of
  Microcalcifications in Mammography}}, Medical Physics 14~(4) (1987) 538--548
  (1987).

\bibitem{Cernadas1998DetectionMC}
E.~Cernadas, R.~Zwiggelaar, W.~Veldkamp, T.~Parr, S.~Astley, C.~Taylor,
  C.~Boggis, {Detection of mammographic microcalcifications using a statistical
  model}, in: N.~Karssemeijer, M.~Thijssen, J.~Hendriks, L.~VanErning (Eds.),
  Digital Mammography, Vol.~13, 1998, pp. 205--208 (1998).

\bibitem{Shi2018HierarchicalPipelineForBreast}
P.~Shi, J.~Zhong, A.~Rampun, H.~Wang, A hierarchical pipeline for breast
  boundary segmentation and calcification detection in mammograms, Computers in
  biology and medicine 96 (2018) 178--188 (2018).

\bibitem{Chan1995MammographicMicrocalcification}
H.~P. Chan, S.~C.~B. Lo, B.~Sahiner, K.~L. Lam, M.~A. Helvie,
  \href{https://doi.org/Doi 10.1118/1.597428}{{Computer-Aided Detection of
  Mammographic Microcalcifications - Pattern-Recognition With an Artificial
  Neural-Network}}, Medical Physics 22~(10) (1995) 1555--1567 (1995).

\bibitem{Gurcan2002OptimalNeuralNetworkArchitecture}
M.~N. Gurcan, H.-P. Chan, B.~Sahiner, L.~Hadjiiski, N.~Petrick, M.~A. Helvie,
  \href{http://dx.doi.org/10.1016/S1076-6332(03)80187-3}{Optimal neural network
  architecture selection}, Academic Radiology 9~(4) (2002) 420–429 (Apr
  2002).

\bibitem{Ge2006ClusterMicrocalcification}
J.~Ge, B.~Sahiner, L.~M. Hadjiiski, H.~P. Chan, J.~Wei, M.~A. Helvie, C.~Zhou,
  \href{https://doi.org/Doi 10.1118/1.2211710}{{Computer Aided Detection of
  Clusters of Microcalcifications on Full Field Digital mammograms}}, Medical
  Physics 33~(8) (2006) 2975--2988 (2006).

\bibitem{Elnaqa2002SVM}
I.~El-Naqa, Y.~Yang, M.~N. Wernick, N.~P. Galatsanos, R.~M. Nishikawa,
  \href{https://doi.org/10.1109/TMI.2002.806569}{{A Support Vector Machine
  Approach for Detection of microcalcifications}}, IEEE Trans Med Imaging
  21~(12) (2002) 1552--1563 (2002).

\bibitem{Wei2005RVM}
L.~Wei, Y.~Yang, R.~M. Nishikawa, M.~N. Wernick, A.~Edwards,
  \href{https://doi.org/10.1109/TMI.2005.855435}{{Relevance Vector Machine for
  Automatic Detection of Clustered microcalcifications}}, IEEE Trans Med
  Imaging 24~(10) (2005) 1278--1285 (2005).

\bibitem{Chan1990ImproveCAD}
H.~P. Chan, K.~Doi, C.~J. Vyborny, R.~A. Schmidt, C.~E. Metz, K.~L. Lam,
  T.~Ogura, Y.~Z. Wu, H.~Macmahon, {Improvement in Radiologists Detection of
  Clustered Microcalcifications on Mammograms - the Potential of Computer-Aided
  Diagnosis}, Investigative Radiology 25~(10) (1990) 1102--1110 (1990).

\bibitem{Chan1999Improvements}
H.~P. Chan, B.~Sahiner, M.~A. Helvie, N.~Petrick, M.~A. Roubidoux, T.~E.
  Wilson, D.~D. Adler, C.~Paramagul, J.~S. Newman, S.~Sanjay-Gopal,
  {Improvement of Radiologists' Characterization of Mammographic Masses By
  Using Computer-Aided Diagnosis: an Roc study}, Radiology 212~(3) (1999)
  817--827 (1999).

\bibitem{Jiang1999ImprovingBreast}
Y.~Jiang, R.~M. Nishikawa, R.~A. Schmidt, C.~E. Metz, M.~L. Giger, K.~Doi,
  {Improving Breast Cancer Diagnosis With Computer-Aided diagnosis}, Academic
  Radiology 6~(1) (1999) 22--33 (1999).

\bibitem{Freer2001ScreeningMammography}
T.~W. Freer, M.~J. Ulissey, {Screening Mammography With Computer-Aided
  Detection: Prospective Study of 12,860 Patients in a Community Breast
  center}, Radiology 220~(3) (2001) 781--786 (2001).

\bibitem{Petrick2002EvaluationMass}
N.~Petrick, B.~Sahiner, H.~P. Chan, M.~A. Helvie, S.~Paquerault, L.~M.
  Hadjiiski, {Breast Cancer Detection: Evaluation of a Mass-Detection Algorithm
  for Computer-Aided Diagnosis -- Experience in 263 patients}, Radiology
  224~(1) (2002) 217--224 (2002).

\bibitem{Hadjiiski2004Improvements}
L.~Hadjiiski, H.~P. Chan, B.~Sahiner, M.~A. Helvie, M.~A. Roubidoux, C.~Blane,
  C.~Paramagul, N.~Petrick, J.~Bailey, K.~Klein, M.~Foster, S.~Patterson,
  D.~Adler, A.~Nees, J.~Shen, \href{https://doi.org/DOI
  10.1148/radiol.2331030432}{{Improvement in Radiologists' Characterization of
  Malignant and Benign Breast Masses on Serial Mammograms With Computer-Aided
  Diagnosis: an Roc study}}, Radiology 233~(1) (2004) 255--265 (2004).

\bibitem{Ko2006ProspectiveAssCAD}
J.~M. Ko, M.~J. Nicholas, J.~B. Mendel, P.~J. Slanetz,
  \href{https://doi.org/187/6/1483 [pii] 10.2214/AJR.05.1582}{{Prospective
  Assessment of Computer-Aided Detection in Interpretation of Screening
  mammography}}, AJR Am J Roentgenol 187~(6) (2006) 1483--1491 (2006).

\bibitem{Gilbert2008SingleReading}
F.~J. Gilbert, S.~M. Astley, M.~G. Gillan, O.~F. Agbaje, M.~G. Wallis,
  J.~James, C.~R. Boggis, S.~W. Duffy, \href{https://doi.org/NEJMoa0803545
  [pii] 10.1056/NEJMoa0803545}{{Single Reading With Computer-Aided Detection
  for Screening mammography}}, N Engl J Med 359~(16) (2008) 1675--1684 (2008).

\bibitem{Gur2004ChangesIn}
D.~Gur, J.~H. Sumkin, H.~E. Rockette, M.~Ganott, C.~Hakim, L.~Hardesty, W.~R.
  Poller, R.~Shah, L.~Wallace, {Changes in Breast Cancer Detection and
  Mammography Recall Rates After the Introduction of a Computer-Aided Detection
  system}, J Natl Cancer Inst 96~(3) (2004) 185--190 (2004).

\bibitem{Fenton2007InfCADMammography}
J.~J. Fenton, S.~H. Taplin, P.~A. Carney, L.~Abraham, E.~A. Sickles, C.~D'Orsi,
  E.~A. Berns, G.~Cutter, R.~E. Hendrick, W.~E. Barlow, J.~G. Elmore,
  \href{https://doi.org/10.1056/NEJMoa066099}{{Influence of Computer-Aided
  Detection on Performance of Screening Mammography}}, N Engl J Med 356~(14)
  (2007) 1399--1409 (2007).

\bibitem{Hall2007BreastImagingCAD}
F.~M. Hall, \href{https://doi.org/10.1056/NEJMe078028}{{Breast Imaging and
  Computer-Aided Detection}}, N Engl J Med 356~(14) (2007) 1464--1466 (2007).

\bibitem{Balleyguier2005CADmammography}
C.~Balleyguier, K.~Kinkel, J.~Fermanian, S.~Malan, G.~Djen, P.~Taourel,
  O.~Helenon, \href{https://doi.org/S0720-048X(05)00024-0 [pii]
  10.1016/j.ejrad.2004.11.021}{{Computer-Aided Detection (cad) in Mammography:
  Does It Help the Junior Or the Senior radiologist?}}, Eur J Radiol 54~(1)
  (2005) 90--96 (2005).

\bibitem{Harris1997b}
K.~M. Harris, V.~G. Vogel, {Breast Cancer screening}, Cancer and Metastasis
  Reviews 16~(3-4) (1997) 231--262 (1997).

\bibitem{Krupinski2008b}
E.~A. Krupinski, Y.~L. Jiang,
  \href{https://doi.org/10.1118/1.2830376}{{Anniversary Paper: Evaluation of
  Medical Imaging systems}}, Medical Physics 35~(2) (2008) 645--659 (2008).

\bibitem{Malich2003Influence}
A.~Malich, D.~Sauner, C.~Marx, M.~Facius, T.~Boehm, S.~O. Pfleiderer, M.~Fleck,
  W.~A. Kaiser, \href{https://doi.org/10.1148/radiol.2283011906 2283011906
  [pii]}{{Influence of Breast Lesion Size and Histologic Findings on Tumor
  Detection Rate of a Computer-Aided Detection system}}, Radiology 228~(3)
  (2003) 851--856 (2003).

\bibitem{Brem2007ImpactOfBreastDensity}
R.~F. Brem, J.~W. Hoffmeister, J.~A. Rapelyea, G.~Zisman, K.~Mohtashemi,
  G.~Jindal, M.~P. Disimio, S.~K. Rogers, \href{https://doi.org/184/2/439
  [pii]}{{Impact of Breast Density on Computer-Aided Detection for Breast
  cancer}}, AJR Am J Roentgenol 184~(2) (2005) 439--444 (2005).

\bibitem{Soo2005CADamorphusCalc}
M.~S. Soo, E.~L. Rosen, J.~Q. Xia, S.~Ghate, J.~A. Baker,
  \href{https://doi.org/184/3/887 [pii]}{{Computer-Aided Detection of Amorphous
  calcifications}}, AJR Am J Roentgenol 184~(3) (2005) 887--892 (2005).

\bibitem{Hall2006CADamorphousCalc}
F.~M. Hall, \href{https://doi.org/186/3/902 [pii]
  10.2214/AJR.06.5013}{{Computer-Aided Detection (cad) of Amorphous
  calcifications}}, AJR Am J Roentgenol 186~(3) (2006) 902; author reply 902--3
  (2006).

\bibitem{GarciaManso2013ConsistentPerformance}
A.~Garcia-Manso, C.~J. Garcia-Orellana, H.~Gonzalez-Velasco,
  R.~Gallardo-Caballero, M.~M. Macias,
  \href{https://doi.org/10.1186/1475-925X-12-2}{{Consistent Performance
  Measurement of a System To Detect Masses in Mammograms Based on Blind Feature
  extraction}}, Biomed Eng Online 12 (2013) 2 (2013).

\bibitem{Kallergi1999EvaluatingPerf}
M.~Kallergi, G.~M. Carney, J.~Gaviria, {Evaluating the Performance of Detection
  Algorithms in Digital mammography}, Med Phys 26~(2) (1999) 267--275 (1999).

\bibitem{Baldi2000AssessingAccuracy}
P.~Baldi, S.~Brunak, Y.~Chauvin, C.~A. Andersen, H.~Nielsen, {Assessing the
  Accuracy of Prediction Algorithms for Classification: an overview},
  Bioinformatics 16~(5) (2000) 412--424 (2000).

\bibitem{Chakraborty2004ObserverStudies}
D.~P. Chakraborty, K.~S. Berbaum, \href{https://doi.org/Doi
  10.1118/1.1769352}{{Observer Studies Involving Detection and Localization:
  Modeling, Analysis, and validation}}, Medical Physics 31~(8) (2004)
  2313--2330 (2004).

\bibitem{Wagner2007AssMedImgTutorial}
R.~F. Wagner, C.~E. Metz, G.~Campbell, {Assessment of Medical Imaging Systems
  and Computer Aids: a Tutorial Review}, Academic Radiology 14 (2007) 723--748
  (2007).

\bibitem{Yoon2007Evaluating}
H.~J. Yoon, B.~Zheng, B.~Sahiner, D.~P. Chakraborty, {Evaluating Computer-Aided
  Detection algorithms}, Med Phys 34~(6) (2007) 2024--2038 (2007).

\bibitem{He2009AlphabetSoup}
X.~He, E.~Frey, \href{https://doi.org/10.1016/j.jacr.2009.06.001}{{ROC, Lroc,
  Froc, Afroc: an Alphabet soup}}, J Am Coll Radiol 6~(9) (2009) 652--655
  (2009).

\bibitem{Wunderlich2012NonparametricLROC}
A.~Wunderlich, F.~Noo, \href{https://doi.org/Doi 10.1109/Tmi.2012.2205015}{{A
  Nonparametric Procedure for Comparing the Areas Under Correlated Lroc
  Curves}}, IEEE Transactions on Medical Imaging 31~(11) (2012) 2050--2061
  (2012).

\bibitem{Yousef2004ComparisonOf}
W.~A. Yousef, R.~F. Wagner, M.~H. Loew, {Comparison of Non-Parametric Methods
  for Assessing Classifier Performance in Terms of {\{}ROC{\}} Parameters}, in:
  Applied Imagery Pattern Recognition Workshop, 2004. Proceedings. 33rd; IEEE
  Computer Society, 2004, pp. 190--195 (2004).

\bibitem{Yousef2006AssessClass}
W.~A. Yousef, R.~F. Wagner, M.~H. Loew, {Assessing Classifiers From Two
  Independent Data Sets Using {\{}ROC{\}} Analysis: a Nonparametric Approach},
  Pattern Analysis and Machine Intelligence, IEEE Transactions on 28~(11)
  (2006) 1809--1817 (2006).

\bibitem{Yousef2019AUCSmoothness-arxiv}
W.~A. Yousef, {AUC}: nonparametric estiamtors and their smoothness, arXiv
  preprint arXiv:1907.12851 (2019).

\bibitem{Yousef2019PrudenceWhenAssumingNormality-arxiv}
W.~A. Yousef, Prudence when assuming normality: an advice for machine learning
  practitioners, arXiv preprint arXiv:1907.12852 (2019).

\bibitem{Fukunaga1984OptimaGlobalNN}
K.~Fukunaga, T.~E. Flick, {AN Optimal Global Nearest Neighbor METRIC}, Ieee
  Transactions on Pattern Analysis and Machine Intelligence 6~(3) (1984)
  314--318 (1984).

\bibitem{Short1981OptimalMeasureNN}
R.~D. Short, K.~Fukunaga, \href{https://doi.org/10.1109/TIT.1981.1056403}{{The
  Optimal Distance Measure for Nearest Neighbor classification}}, Information
  Theory, IEEE Transactions on 27~(5) (1981) 622--627 (1981).

\bibitem{Elsayed2019Matlab-arxiv}
A.~A. Elsayed, W.~A. Yousef, Matlab vs. opencv: A comparative study of
  different machine learning algorithms, arXiv preprint arXiv:1905.01213
  (2019).

\bibitem{LIBCADUtil2012}
{MESC Labs}, {LIBCAD-OpenSource: Software Utilities for Computer Aided
  Detection (CAD)}: {https://github.com/mesclabs/LIBCAD-OpenSource-Utils}
  (2012).

\bibitem{Sacred2017}
K.~Greff, A.~Klein, M.~Chovanec, F.~Hutter, J.~Schmidhuber, The sacred
  infrastructure for computational research: {https://github.com/IDSIA/sacred},
  Proceedings of the 15th Python in Science Conference (2017) 49--56 (2017).

\bibitem{Kim2010ComparisonCommercialCAD}
S.~J. Kim, W.~K. Moon, S.~Y. Kim, J.~M. Chang, S.~M. Kim, N.~Cho,
  \href{https://doi.org/10.3109/02841851003709490}{{Comparison of Two Software
  Versions of a Commercially Available Computer-Aided Detection (cad) System
  for Detecting Breast cancer}}, Acta Radiol 51~(5) (2010) 482--490 (2010).

\bibitem{Roehrig2005ManufacturerePresp}
J.~Roehrig, \href{https://doi.org/78/suppl_1/S41 [pii]
  10.1259/bjr/25058162}{{The Manufacturer's perspective}}, Br J Radiol 78 Spec
  No (2005) S41--5 (2005).

\bibitem{Ellis2007EvaluationCAD}
R.~L. Ellis, A.~A. Meade, M.~A. Mathiason, K.~M. Willison, W.~Logan-Young,
  \href{https://doi.org/10.1148/radiol.2451060760}{{Evaluation of
  Computer-Aided Detection Systems in the Detection of Small Invasive Breast
  Carcinoma1}}, Radiology 245~(1) (2007) 88--94 (2007).

\bibitem{Brem2005CAD}
R.~F. Brem, J.~W. Hoffmeister, G.~Zisman, M.~P. DeSimio, S.~K. Rogers,
  \href{https://doi.org/184/3/893 [pii]}{{A Computer-Aided Detection System for
  the Evaluation of Breast Cancer By Mammographic Appearance and Lesion size}},
  AJR Am J Roentgenol 184~(3) (2005) 893--896 (2005).

\bibitem{Lobbes2013MalignantLesions}
M.~Lobbes, M.~Smidt, K.~Keymeulen, R.~Girometti, C.~Zuiani, R.~Beets-Tan,
  J.~Wildberger, C.~Boetes,
  \href{https://doi.org/10.1016/j.clinimag.2012.04.017}{{Malignant Lesions on
  Mammography: Accuracy of Two Different Computer-Aided Detection systems}},
  Clinical Imaging 37~(2) (2013) 283--288 (2013).

\bibitem{Kodak2004}
Kodak, {Kodak Mammography, CAD Engine - P030007} (2004).

\bibitem{Kuroki2012PerformanceEval}
Y.~Kuroki, R.~Sekiguchi, T.~Endo, Utsonomiya/JP, T.~Utsunomiya, Nagoya/JP,
  \href{https://doi.org/10.1594/ecr2012/C-0930}{{FujifilmCAD.pdf}}, in:
  Performance evaluation of the mammography CAD (Computer Aided Detection) on
  the direct conversion FPD system with a pixel size of 50$\mu$m, ECR 2012 /
  C-0930, 2012 (2012).

\bibitem{Carneiro2017DeepLearningModelsFor}
G.~Carneiro, J.~Nascimento, A.~P. Bradley, Deep learning models for classifying
  mammogram exams containing unregistered multi-view images and segmentation
  maps of lesions, in: Deep Learning for Medical Image Analysis, Elsevier,
  2017, pp. 321--339 (2017).

\bibitem{Akselrod-Ballin2017DeepLearningForAutomatic}
A.~Akselrod-Ballin, L.~Karlinsky, A.~Hazan, R.~Bakalo, A.~B. Horesh,
  Y.~Shoshan, E.~Barkan, Deep learning for automatic detection of abnormal
  findings in breast mammography, in: Deep Learning in Medical Image Analysis
  and Multimodal Learning for Clinical Decision Support, Springer, 2017, pp.
  321--329 (2017).

\bibitem{Akselrod-Ballin2017CnnBasedMethodFor}
A.~Akselrod-Ballin, L.~Karlinsky, S.~Alpert, S.~Hashoul, R.~Ben-Ari, E.~Barkan,
  A cnn based method for automatic mass detection and classification in
  mammograms, Computer Methods in Biomechanics and Biomedical Engineering:
  Imaging \& Visualization (2017) 1--8 (2017).

\bibitem{Qiu2017NewApproachDevelopComputer}
Y.~Qiu, S.~Yan, R.~R. Gundreddy, Y.~Wang, S.~Cheng, H.~Liu, B.~Zheng, A new
  approach to develop computer-aided diagnosis scheme of breast mass
  classification using deep learning technology, Journal of X-ray science and
  technology 25~(5) (2017) 751--763 (2017).

\bibitem{Hamidinekoo2018DeepLearningMammographyBreast}
A.~Hamidinekoo, E.~Denton, A.~Rampun, K.~Honnor, R.~Zwiggelaar, Deep learning
  in mammography and breast histology, an overview and future trends, Medical
  image analysis 47 (2018) 45--67 (2018).

\bibitem{Dhungel2017DeepLearningApproachFor}
N.~Dhungel, G.~Carneiro, A.~P. Bradley, A deep learning approach for the
  analysis of masses in mammograms with minimal user intervention, Medical
  image analysis 37 (2017) 114--128 (2017).

\bibitem{Carneiro2017AutomatedAnalysisUnregisteredMulti}
G.~Carneiro, J.~Nascimento, A.~P. Bradley, Automated analysis of unregistered
  multi-view mammograms with deep learning, IEEE transactions on medical
  imaging 36~(11) (2017) 2355--2365 (2017).

\bibitem{Dhungel2017FullyAutomatedClassificationMammograms}
N.~Dhungel, G.~Carneiro, A.~P. Bradley, Fully automated classification of
  mammograms using deep residual neural networks, in: Biomedical Imaging (ISBI
  2017), 2017 IEEE 14th International Symposium on, IEEE, 2017, pp. 310--314
  (2017).

\bibitem{Bevilacqua2018PerformanceComparisonBetweenShallow}
V.~Bevilacqua, A.~Brunetti, A.~Guerriero, G.~F. Trotta, M.~Telegrafo,
  M.~Moschetta, A performance comparison between shallow and deeper neural
  networks supervised classification of tomosynthesis breast lesions images,
  Cognitive Systems Research (2018).

\bibitem{Tourassi2006StudyOnComp}
G.~D. Tourassi, D.~M. Delong, C.~E. Floyd,
  \href{https://doi.org/10.1088/0031-9155/51/5/018}{{A Study on the
  Computerized Fractal Analysis of Architectural Distortion in Screening
  mammograms}}, Physics in Medicine and Biology 51~(5) (2006) 1299--1312
  (2006).

\bibitem{Sampat2005EvidenceArchDist}
M.~P. Sampat, G.~J. Whitman, M.~K. Markey, A.~C. Bovik,
  \href{https://doi.org/10.1117/12.595331}{{Evidence Based Detection of
  Spiculated Masses and Architectural distortions}}, Medical Imaging 2005:
  Image Processing, Pt 1-3 5747 (2005) 26--37 (2005).

\bibitem{Matsubara2003AutoDetectMethod}
T.~Matsubara, T.~Ichikawa, T.~Hara, H.~Fujita, S.~Kasai, T.~Endo, T.~Iwase,
  \href{https://doi.org/10.1016/S0531-5131(03)00496-5}{{Automated Detection
  Methods for Architectural Distortions Around Skinline and Within Mammary
  Gland on mammograms}}, Cars 2003: Computer Assisted Radiology and Surgery,
  Proceedings 1256 (2003) 950--955 (2003).

\bibitem{Ichikawa2004AutoDetect}
T.~Ichikawa, T.~Matsubara, T.~Hara, H.~Fujita, T.~Endo, T.~Iwase,
  \href{https://doi.org/10.1117/12.535116}{{Automated Detection Method for
  Architectural Distortion Areas on Mammograms Based on Morphological
  Processing and Surface analysis}}, Medical Imaging 2004: Image Processing,
  Pts 1-3 5370 (2004) 920--925 (2004).

\bibitem{Guo2005InvestSVM}
Q.~Guo, J.~Shao, V.~Ruiz,
  \href{https://doi.org/10.1088/1742-6596/15/1/015}{{Investigation of Support
  Vector Machine for the Detection of Architectural Distortion in Mammographic
  images}}, Sensors {\&} Their Applications XIII 15 (2005) 88--94 (2005).

\bibitem{Ayres2005CharacArchDist}
F.~J. Ayres, R.~M. Rangayyan, {Characterization of Architectural Distortion in
  mammograms}, Ieee Engineering in Medicine and Biology Magazine 24~(1) (2005)
  59--67 (2005).

\bibitem{Celaya-Padilla2018ContralateralAsymmetryForBreast}
J.~M. Celaya-Padilla, C.~H. Guzm{\'a}n-Valdivia, C.~E. Galv{\'a}n-Tejada, J.~I.
  Galv{\'a}n-Tejada, H.~Gamboa-Rosales, I.~Garza-Veloz, M.~L. Martinez-Fierro,
  M.~A. Cid-B{\'a}ez, A.~Martinez-Torteya, F.~J. Martinez-Ruiz, et~al.,
  Contralateral asymmetry for breast cancer detection: a cadx approach,
  Biocybernetics and Biomedical Engineering 38~(1) (2018) 115--125 (2018).

\bibitem{Yin1991ComputerizedDetection}
F.-F. Yin, M.~L. Giger, K.~Doi, C.~E. Metz, C.~J. Vyborny, R.~A. Schmidt,
  {Computerized Detection of Masses in Digital Mammograms: Analysis of
  Bilateral Subtraction images}, Medical Physics 18~(5) (1991) 955--963 (1991).

\bibitem{Yin1993ComparisonBilateral}
F.~F. Yin, M.~L. Giger, C.~J. Vyborny, K.~Doi, R.~A. Schmidt, {Comparison of
  Bilateral-Subtraction and Single-Image Processing Techniques in the
  Computerized Detection of Mammographic Masses}, Investigative Radiology
  28~(6) (1993) 473--481 (1993).

\bibitem{Yin1994CompDetectionMasses}
F.~F. Yin, M.~L. Giger, K.~Doi, C.~J. Vyborny, R.~A. Schmidt, {Computerized
  Detection of Masses in Digital Mammograms: Investigation of Feature-Analysis
  techniques.}, J Digital Imaging 7~(1) (1994) 18--26 (1994).

\bibitem{Yin1994ComputerizedAutomated}
F.~F. Yin, M.~L. Giger, K.~Doi, C.~J. Vyborny, R.~A. Schmidt, {COMPUTERIZED
  Detection of Masses in Digital Mammograms - Automated Alignment of Breast
  Images and Its Effect on Bilateral-Subtraction TECHNIQUE}, Medical Physics
  21~(3) (1994) 445--452 (1994).

\bibitem{Scutt2006BreastAsym}
D.~Scutt, G.~A. Lancaster, J.~{T Manning},
  \href{https://doi.org/10.1186/bcr1388}{{Breast Asymmetry and Predisposition
  To Breast cancer}}, Breast Cancer Research 8~(2) (2006).

\bibitem{Rangayyan2007AnaBilateralAsym}
R.~M. Rangayyan, R.~J. Ferrari, A.~F. Frere,
  \href{https://doi.org/10.1117/1.2712461}{{Analysis of Bilateral Asymmetry in
  Mammograms Using Directional, Morphological, and Density features}}, Journal
  of Electronic Imaging 16~(1) (2007).

\bibitem{Lau1991AutoDet}
T.~K. Lau, W.~F. Bischof, {Automated Detection of Breast-Tumors Using the
  Asymmetry Approach}, Computers and Biomedical Research 24~(3) (1991) 273--295
  (1991).

\bibitem{Ferrari2001AnaAsym}
R.~J. Ferrari, R.~M. Rangayyan, J.~E.~L. Desautels, A.~F. Frere, {Analysis of
  Asymmetry in Mammograms Via Directional Filtering With Gabor wavelets}, IEEE
  Transactions on Medical Imaging 20~(9) (2001) 953--964 (2001).

\end{thebibliography}
